\documentclass[english,notitlepage, hyperref]{aastex6}
\usepackage[T1]{fontenc}
\usepackage[latin9]{inputenc}
\usepackage{wrapfig}
\usepackage{amsbsy}
\usepackage{amstext}
\usepackage{amssymb}
\usepackage{graphicx}

\makeatletter

\providecommand{\tabularnewline}{\\}

\def\aap{{Astronomy and Astrophys.}}

\def\apj{{\it Astrophys. J.\ }}
\def\apjs{{\apj\ \it Suppl.\ }}		
\def\apjl{{\apj\ \it Lett.\ }}

\usepackage{hyperref}
\renewcommand{\baselinestretch}{1.0}
\makeatother

\usepackage{babel}
\begin{document}

\title{Cosmic Ray Spectrum Steepening in Supernova Remnants - I. Loss-Free
Self-Similar Solution}

\author{M.A. Malkov}

\affil{Department of Physics and CASS, University of California San Diego}

\author{F.A. Aharonian}

\affil{Dublin Institute for Advanced Studies, 31 Fitzwilliam Place, Dublin
2, Ireland}

\affil{Max-Planck-Institut für Kernphysik, P.O. Box 103980, D 69029 Heidelberg,
Germany}
\begin{abstract}
The direct measurements of cosmic rays (CRs), after correction for
the propagation effects in the interstellar medium, indicate that
their source spectra are likely to be significantly steeper than the
canonical $E^{-2}$ spectrum predicted by the standard Diffusive Shock
Acceleration (DSA) mechanism. The DSA has long been held responsible
for the production of galactic CRs in supernova remnant (SNR) shocks.
The $\gamma$-ray \textquotedbl probes\textquotedbl{} of the acceleration
spectra of CRs on-the-spot, inside of the SNRs, lead to the same conclusion.
We show that the steep acceleration spectrum can be attributed to
the \emph{combination} of (i) spherical expansion, (ii) tilting of
the magnetic field along the shock surface and (iii) shock deceleration.
Because of (i) and (ii), the DSA is efficient only on two ``polar
caps'' of a spherical shock where the local magnetic field is within
$\simeq45^{\circ}$ to its normal. The shock-produced spectrum observed
edge-on steepens with the particle energy because the number of freshly
accelerated particles with lower energies continually adds up to a
growing acceleration region. We demonstrate the steepening effect
by obtaining an exact self-similar solution for the particle acceleration
at expanding shock surface with an arbitrary energy dependence of
particle diffusivity $\kappa$. We show that its increase toward higher
energy steepens the spectrum, which deeply contrasts with the standard
DSA spectrum where $\kappa$ cancels out.
\end{abstract}

\section{Introduction}

It is believed that the relativistic particles, electrons, protons
and nuclei, measured locally in the Solar System, represent, at least
up to $E\sim10^{15}$eV, the so-called Galactic component of CRs.
The flux of these particles is determined by the overall rate of injection
of CRs into the interstellar medium (ISM) and by their confinement
time in the Galactic Disk, $\tau$. Because of the energy-dependent
propagation, the energy spectrum of CRs in the Galactic Disk is modified
and significantly deviates from the average source (acceleration)
spectrum. The latter is recovered from the energy distribution of
primary cosmic rays $dN/dE\propto E^{-2.75-2.85}$, and the energy
dependence of the confinement $\tau(E)\propto E^{-\delta}$, with
$\delta\sim0.3-0.5$ as it follows from the independently reported
CR secondary-to-primary ratio. The resulting mismatch between the
source spectrum, ought to be in the range $E^{-2.25}-E^{-2.55}$,
and the \textquotedbl nominal\textquotedbl{} $E^{-2}$ type spectrum
predicted by the standard Diffusive Shock Acceleration (DSA) theory,
can no longer be attributed to the statistical or systematic errors
in observations. Indeed, the latter now reliably produce CR spectral
indices with up to three-digit precision in relevant energy ranges.
Moreover, it is becoming increasingly evident that $\gamma$-ray observations,
which provide a powerful tool for the \emph{in situ} probes of energy
distributions of CRs inside the accelerators, also point to a similar
discrepancy. Despite the DSA predictions, most of SNRs, including
very young objects like SN 1006, Tycho, and Cas A, show fairly steep
spectra \citep{Aharon2018}. Instead of a single $E^{-2}$ -type spectrum
stretching all the way to the maximum energy, which would serve as
a signature of the DSA, we often see softer spectra showing a gradual
steepening. This descrepancy is a growing concern in the CR community,
reflected, e.g., in a recent review by \citealp{Gabici2019} 
\footnote{Short after our manuscript was accepted for publication, 
a new paper addressing this same issue was posted, \citep{Bell2019}}.

Logically, such disagreement may be attempted to settle by tweaking
the DSA theory. However, the DSA is remarkably insensetive to factors
other than the shock compression as long as no strong backreaction
of accelerated particles on the shock itself is expected. To make
matters worse, such backreaction generally leads to harder rather
than softer spectra for energies exceeding several GeV. Indeed, an
analytic solution incorporating the backreaction has a\emph{concave
}spectrum tending to $E^{-3/2}$ at high energies, in the limit of
strong shocks and maximum energy in the TeV range \citep{MDru01}.
This trend is clearly in conflict with the \emph{convex }spectra inferred
from the $\gamma$-emission in a number of SNRs. We will return to
the nonlinear shock modification and some other aspects of the DSA,
which can potentially alter the $E^{-2}$ spectrum, in the Discussion
section.

It follows that the above constraints on the acceleration spectra
may point to some DSA physics not included in standard treatments.
In particular, we will demonstrate that the spectral index $q$, and
the maximum (cutoff) momentum, $p_{{\rm max}}$, are not exhaustive
parameters for the DSA theory, under acceleration conditions outlined
below and explained further in the paper. Even if $p_{{\rm max}}$
can be regarded as adjustable, given the limited acceleration time
and shock size, the power-law index $q$ is a robust DSA prescription:
$f_{CR}\propto p^{-q}$ (we will use $p$ instead of \emph{E}, so
that $p^{-q}$ corresponds to $E^{2-q}$). The index $q$ merely depends
on the Mach number, $M$, but only weakly so if the shock is strong,
$q\approx4/\left(1-M^{-2}\right)$; note that the bulk of the galactic
CRs are thought to come namely from strong shocks. Most of the DSA
treatments, however, apply to planar, slowly evolving (over the acceleration
timescale) shocks. Although spherically expanding shocks have also
been studied, they were approximated by isotropic, essentially 1D
(radial) configurations, e.g., \citep{PtusPrish1981,Ber94DruSuppr,DruryEscape11}
(see, however, \citet{VoelkInj03,Pfrommer2018MNRAS}). The overall
production spectrum in such shocks remains similar to the planar DSA
predictions, with a possible exception of the work by \citet{DruryEscape11}
whose phenomenological ``box'' model has the potential for spectral
steepening. However, the box model incurs the loss of information
due to spatial integration, so the obtained particle spectrum contains
undetermined parameters.

This paper is an attempt to remodel the DSA for capturing three-dimensional
aspects of shock interaction with the ambient magnetic field. Assuming
spherical shock geometry and a homogeneous ambient magnetic field,
one may see that particle acceleration conditions change on any given
field line crossing the shock surface. This is because the angle $\vartheta_{nB}$
that the field line makes with the shock normal decreases with time.
Particle injection (and to some extent their subsequent acceleration)
becomes efficient when $\vartheta_{nB}$ decreases below some critical
value, $\vartheta_{nB}\lesssim\vartheta_{cr}\approx\pi/4$. Our objective
is to understand the impact of this continuous transition on the integrated
CR spectra and, by implication, on the instantaneous line-of-sight
ones. Besides the $\vartheta_{nB}$ variation, we also include the
effects of shock expansion and deceleration. The time dependence of
the DSA comes then in two flavors: the shock's slow down and continuous
addition of freshly accelerated (low-energy) particles to those field
lines that just began to make its angle $\vartheta_{nB}\lesssim\pi/4$.

Effects of magnetic field inclination and time-dependence make the
DSA spectrum \emph{convex}, which we will demonstrate using a closed-form
self-similar solution for an SNR expanding into a homogeneous magnetic
field. Integrating the convection-diffusion equation across the field
makes the problem effectively one-dimensional, but strongly time-dependent
with the 3D effect of magnetic field inclination included. As \emph{particle
losses are neglected}, this solution produces a minimum steepening
effect, caused exclusively by the \emph{momentum dependence of particle
diffusion}. By contrast, the loss-free DSA solution for a steadily
propagating planar shock is fundamentally independent of the particle
diffusion rate because it cancels out from the balance of spatial
and momentum transport terms (see also Sec.\ref{sec:Comparison-with-Standard}
below).

While not addressing the particle losses directly, this paper sets
out a framework for studying an additional spectrum steepening caused
by them. The losses may be incurred in the following ways:
\begin{itemize}
\item direct escape of particles from the acceleration zone along the field
lines by reaching its boundary where both the turbulence and particle
self-confinement are weak, \citep{MetalEsc13,GabiciAnisEsc12,DAngeloBlasi2016}
\item crossfield particle diffusion toward the edge of acceleration zone
($\vartheta_{nB}\approx\vartheta_{cr})$ with a subsequent rapid escape
along the field lines due to insufficient particle self-confinement
in the region $\vartheta\gtrsim\vartheta_{cr}$ (this scenario is
briefly considered in the paper and studied in detail by Hanusch et
al., prepared for submission to ApJ)
\item crossfield expansion of the acceleration zone driven by the pressure
of accelerated particles and possibly amplified magnetic field
\end{itemize}
The focus of this paper is on a probably not very large subset of
SNRs with a bilateral morphology, best exemplified by the SNR 1006.
At the same time, this type of SNRs, and especially the 1006, is an
excellent laboratory to study the DSA, as it links the magnetic field
direction with the acceleration efficiency all around the SNR shell.
Particle acceleration in more complex SNRs with a less regular ambient
field can hardly be understood without a grasp of the DSA operation
in simple bilateral remnants. Indeed, turbulent ISM field with sufficiently
large correlation length being intersected by a shock front (even
if it is planar) will exhibit local alternation between injection
and non-injection shock areas, similar to that observed in SNR 1006
at the scale of the entire remnant. We will also return to such possibility
in the Discussion section. The paper is structured as follows: Sec.\ref{sec:DSA-Mechanism-With}
discusses modifications to the DSA theory for a spherical shock propagation
in a constant magnetic field. In Sec.\ref{sec:Equation-for-Particle}
a convection-diffusion equation for particle acceleration is introduced,
along with some simplifications of the shock geometry, whereas its
solution is obtained in Sec.\ref{sec:Solutions}. After reviewing
its limiting cases in Sec.\ref{sec:Spectral-properties-and} and comparing
the results with those of the standard DSA in Sec.\ref{sec:Comparison-with-Standard},
including particle escape, the paper concludes with a brief summary
and discussion in Secs.\ref{sec:Conclusions} and \ref{sec:Discussion}.

\section{DSA Mechanism With Evolving Shock Obliquity\label{sec:DSA-Mechanism-With}}

\subsection{Particle Acceleration Domain on the Shock Surface}

\begin{figure}
\includegraphics[bb=30bp 60bp 420bp 300bp,scale=0.70]{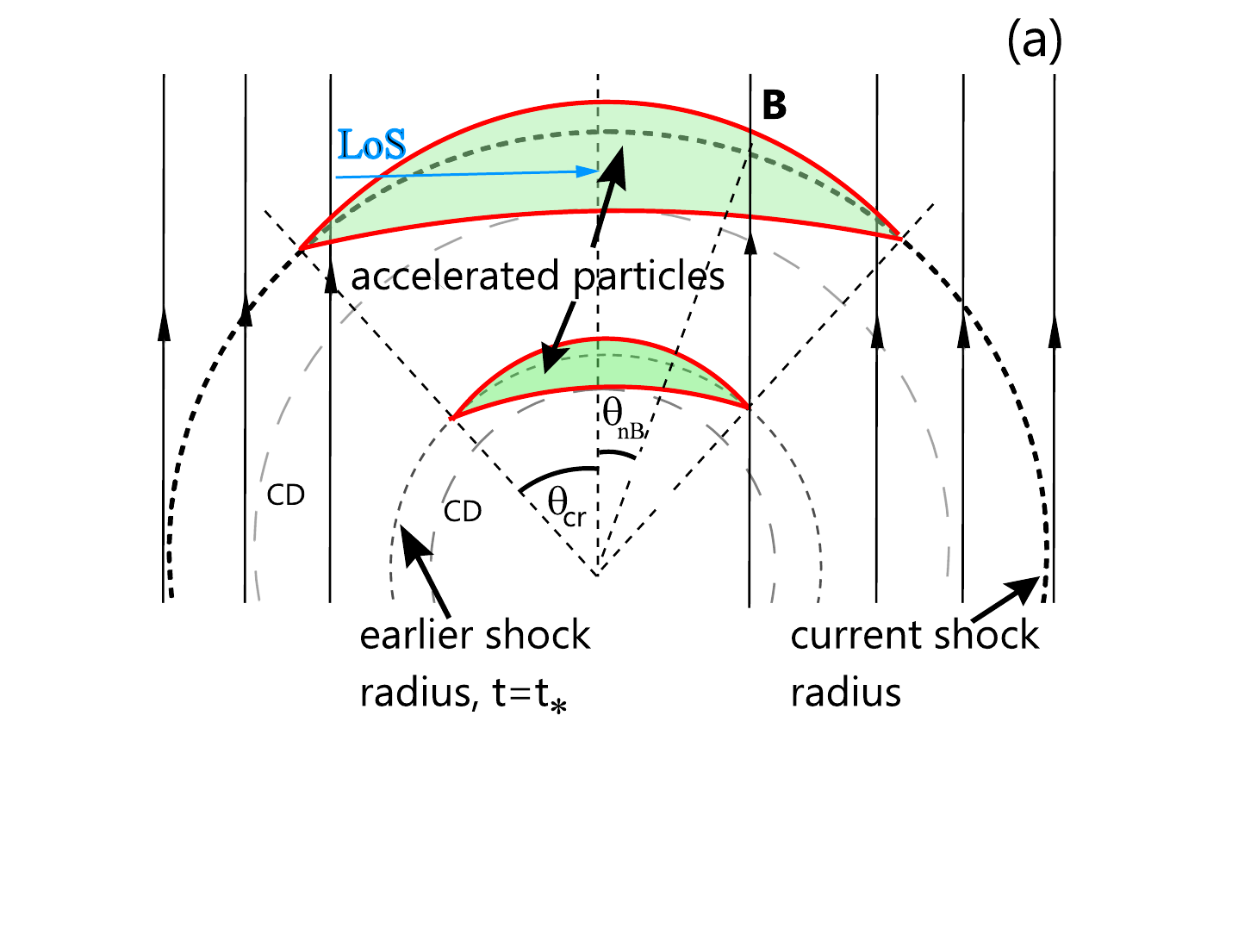}\includegraphics[scale=0.28]{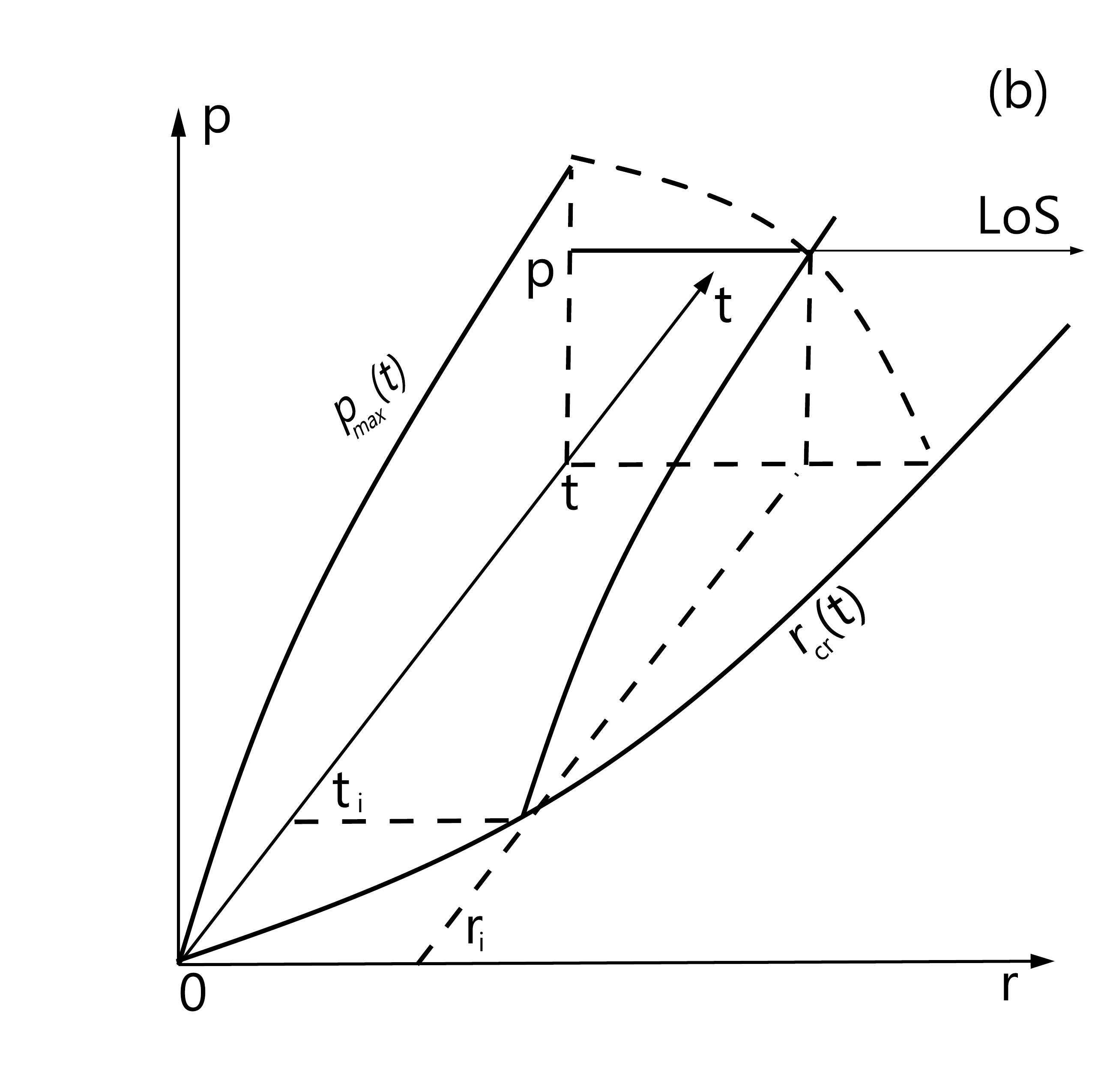}

\caption{(a) A half of the spherical shock surface expanding into an ISM plasma
with a constant magnetic field. The shock surface is shown at two
different times. The long-dashed line circles indicate the respective
positions of the contact discontinuity, CD. An opposite side of the
remnant with a symmetric shock-field alignment is not shown. (b) CR
production in expanding active acceleration zone $r\le r_{\text{cr}}\left(t\right)$,
where $r$ is counted from the magnetic axis $\vartheta=0$ along
the shock front. Particles started their acceleration at $t=t_{i}$,
$r=r_{i}$ do not contribute to the spectrum between $p$ and $p_{\text{max}}\left(t\right)$
by the time $t$. This makes the spectrum integrated along the LoS
softer than the local spectrum at any fixed $r<r_{\text{cr}}\left(t\right)\equiv R_{\text{s}}\left(t\right)\sin\vartheta_{\text{cr}},$
since the length of the LoS (heavy-line portion of it) is becoming
progressively shorter with growing particle momentum.\label{fig:Spherical-shock-surface}}
\end{figure}

The spectral softening mechanism considered in this paper is illustrated
in Fig.\ref{fig:Spherical-shock-surface}. The two crescent regions
in Fig.\ref{fig:Spherical-shock-surface}a show that part of a spherical
shock where the DSA works efficiently. This active acceleration zone
is presented at two separate times to show in Sec.\ref{subsec:Spectrum-Steepening-Mechanism}
below how the spectrum sampled along the line of sight may appear
steeper than the one at any given field line. Shock-accelerated particles
are concentrated at this expanding region because the proton injection
is efficient only where the shock normal makes a reasonably sharp
angle, $\vartheta_{nB}$, with the local magnetic field direction
(quasi-parallel shock geometry), assumed vertical in the figure. The
critical angle $\vartheta_{nB}=\vartheta_{{\rm cr}},$ beyond which
the proton injection rate into the DSA drops sharply, is close to
$\vartheta_{{\rm cr}}\simeq\pi/4$. This choice of the source for
particle injection is supported by simple theoretical considerations
\citep{mv95,VoelkInj03}, as well as Monte-Carlo \citep{Ellison1995ApJ}
and hybrid simulations, e.g., \citep{thomas1990two,CaprSpitk14a}.
A simple explanation is that for larger $\vartheta_{nB}$ the recession
of the field line - shock intersection point is too fast for the downstream
particles to return upstream and continue acceleration. Most importantly,
this pattern of particle acceleration is supported by observations
\citep{LongRaymond03}.

While the shock radius increases, the accelerated particles fill up
an expanding disc-like layer near the shock surface intersecting a
cone with a constant opening angle $\approx\vartheta_{{\rm cr}}$.
This layer consists of two parts separated by the shock surface: the
shock precursor, of the size $\kappa\left(p_{{\rm max}}\right)/U_{{\rm sh}}$,
and the post-shock region occupied by the accelerated particles convected
downstream. (Here $\kappa$ is the particle diffusivity and $U_{{\rm sh}}$
is the shock speed.) The thicknesses of these layers also grow in
time; the precursor grows due to an increase in the maximum momentum,
$p_{{\rm max}}\left(t\right)$, under a balance of particle diffusion
away from the shock while the shock is catching them up. It also grows
because of a decreasing shock speed. In the case of Bohm regime, the
particle diffusion coefficient scales linearly with $p\gg mc,$ as
$\kappa\simeq\kappa_{B}\equiv cr_{g}/3$, where $r_{g}$ is the particle
gyro-radius and $c$ is the speed of light. The thickness of the downstream
particle layer is also determined by diffusion and convection. However,
by contrast with the upstream layer they act in the same direction:
the convective growth of the layer, $L_{conv}\sim U_{{\rm sh}}\left(t-t_{i}\right)/\sigma$,
is extended by a trailing sub-layer across which the density of accelerated
particles drops to its background (e.g., zero) level. The particle
transport across this sub-layer (along the field) is self-regulated
by Alfven waves excited by particles escaping downstream from the
shocked plasma. The time $t_{i}=t_{i}\left(\vartheta,R\right)$ signifies
the beginning of acceleration on a given field line crossed by the
shock at the critical angle $\vartheta_{{\rm cr}}.$ The shock compression
$\sigma$ in the above expression for $L_{conv}$ accounts for the
slow down of the upstream flow upon the shock crossing. Time-asymptotically,
when the far downstream side of this particle sub-layer is disconnected
from the shock, a self-similar solution obtained by \citep{MetalEsc13}
applies. In this paper, however, we do not take into account this
latter aspect of particle transport and focus on the anisotropic magnetic
environment, essential for the CR production spectrum but disregarded
in previous studies. The above brief description of particle transport
upstream and downstream of the shock will be useful in constructing
time-dependent self-similar solution.

\subsection{Spectrum Steepening Mechanism\label{subsec:Spectrum-Steepening-Mechanism}}

When the particle acceleration zone near a shock is seen edge-on,
which is typical because of a greatly enhanced emissivity, the line-of-sight
(LoS) integrated emission is sampled from particles with different
acceleration history, depending on their distance from the magnetic
axis ($\vartheta=0$); it is the longest at $\vartheta=0$. Farther
away from the magnetic axis of the remnant, where $\vartheta$ approaches
$\vartheta_{nB}\simeq\vartheta_{{\rm cr}}$, the freshly accelerated
particles contribute to lower momenta of the observed spectrum, Fig.\ref{fig:Spherical-shock-surface}b,
because of the shorter duration of acceleration on the field lines
near $\vartheta_{{\rm cr}}$. They make thus the LoS-integrated spectrum
steeper by enriching its low-energy part.

To demonstrate the spectrum steepening caused by the shock geometry,
consider an idealized case when the freshly injected particles undergo
shock acceleration but remain on the same field line (do not diffuse
along the shock front). This assumption is justified (at least for
$t\lesssim\tau_{a}\sim\kappa_{\parallel}/U_{{\rm sh}}^{2}$) because
the diffusion length along the field line (shock precursor scale)
$L_{p}\sim\kappa_{\parallel}/U_{sh}\ll r_{{\rm cr}}\sim R_{s}$, so
that the lateral displacement during the acceleration time $\tau_{a}\sim$$\kappa_{\parallel}/U_{sh}^{2}$
is $L_{\perp}\sim$$\sqrt{\kappa_{\perp}\tau_{a}}=\sqrt{\kappa_{\perp}\kappa_{\parallel}}/U_{sh}\sim\kappa_{B}/U_{sh}\sim L_{p}\ll r_{{\rm cr}}$.
Apart from the normalization, unimportant here, the spectrum with
an abrupt cutoff at $p_{{\rm max}}$ at any given field line at a
distance $r$ from the axis is

\[
f\left(p,r\right)=p^{-q}H\left(p_{\max}\left(r\right)-p\right)
\]
where $H$ is the Heaviside unit function, and $p_{\max}$ is the
maximum momentum at the distance from magnetic axis, $r\approx R_{s}\sin\vartheta_{nB}$,
Fig.\ref{fig:Spherical-shock-surface}. Assuming Bohm diffusion and
a ballistic shock expansion, which implies $p_{max}\left(r=0,t\right)\propto r_{{\rm cr}}\left(t\right)\propto t$,
we can write $p_{\max}\left(r\right)=p_{\max}\left(0,t\right)\left(1-r/r_{{\rm cr}}\right)$.
We can then define the following spatially averaged spectrum

\[
F_{l}\left(p\right)=\frac{1}{r_{{\rm cr}}^{l+1}}\int_{0}^{r_{\text{cr}}}r^{l}f\left(p,r\right)dr
\]
where $l=\left\{ 0,1\right\} $ for a LoS-averaged ($l=0$) spectrum
or shock-area integrated one ($l=1$). Note that the latter case is
more closely related to the cumulative CR production and will be considered
in detail in the next section. It is also clear that the effect of
spectral steepening caused by addition of freshly injected particles
at $r\approx r_{{\rm cr}}$ is stronger in this case. Substituting
the above expressions for $f$ and $p_{\max},$ and assuming that
(locally) the DSA produces the spectrum $\propto p^{-q}$, with $q=3\sigma/\left(\sigma-1\right)$,
where $\sigma$ is the shock compression ratio, we find
\[
F_{l}\left(p\right)=\frac{p^{-q}}{l+1}\left(1-\frac{p}{p_{\max}\left(0\right)}\right)^{l+1}
\]
For $p\ll p_{\max}\left(0\right)$, the latter result can be written
as

\[
F_{l}\left(p\right)\approx\frac{1}{l+1}p^{-q}\exp\left[-\left(l+1\right)\frac{p}{p_{\max}\left(0\right)}\right]
\]
The last expression shows that the standard DSA spectrum $p^{-q}$
steepens with momentum and can even mimic an incipient exponential
cut-off if observed at moderately high momenta. In the next section
we take a more systematic approach to this effect by including also
the particle cross-field diffusion and time dependent shock evolution.

\section{Equation for Particle Acceleration\label{sec:Equation-for-Particle}}

Since the direction of magnetic field relative to the shock normal
is key for particle injection, it is convenient to use a cylindrical
coordinate system, $\left(r_{\perp},z\right)$, with $\boldsymbol{\mathbf{B}}\parallel\hat{\boldsymbol{\mathbf{z}}}$,
where $\hat{\mathbf{z}}$ is the unit vector along the $z-$ axis
passing through the center of the remnant. The overall shock morphology
in such remnants as SNR 1006 and Tycho is spherical, but at least
the 1006 appearance is distinctively different in the areas of quasi-parallel
and quasi-perpendicular shock geometry ($\vartheta_{nB}\ll1$ and
$\pi/2-\vartheta_{nB}\ll1,$ respectively). In the NE and SW regions
of the 1006, large portions of the shock surface are flattened, if
not slightly concave or at least corrugated, as we can guess from
our vista point, \citep{LongRaymond03,Bamba05,Cassam-Ch2008}. The
overall shock morphology is therefore a barrel-shaped rather than
spherical, with the equatorial part resembling a cylindrical surface
rather than spherical shell. Since the active acceleration zones coincide
with two \emph{flattened} polar caps, we can ascribe two unique values
to the respective shock coordinates, $z=\pm z_{s}\left(t\right)$,
assumed independent of the radial coordinate, $r_{\perp}$. The radial
extent of that portion of the shock surface where the particle acceleration
is efficient is limited by the critical angle $\vartheta_{nB}=\vartheta_{{\rm cr}}\simeq\pi/4$,
which justifies the locally planar shock approximation. The convection-diffusion
equation for particles undergoing acceleration in the layer near $z=z_{s}\left(t\right)$
and $r_{\perp}\leq r_{{\rm cr}}\left(t\right)=R_{s}\left(t\right)\sin\vartheta_{{\rm cr}}$
takes the following form

\begin{equation}
\frac{\partial f}{\partial t}+u\frac{\partial f}{\partial z}-\frac{\partial}{\partial z}\kappa_{\parallel}\frac{\partial f}{\partial z}-\frac{1}{r_{\perp}}\frac{\partial}{\partial r_{\perp}}r_{\perp}\kappa_{\perp}\frac{\partial f}{\partial r_{\perp}}-\frac{1}{3}\frac{\partial u}{\partial z}p\frac{\partial f}{\partial p}=Q\left(r_{\perp},t\right)\delta\left(z\right)\delta\left(p-p_{0}\right)\label{eq:CDeq}
\end{equation}
Here we shifted the origin of $z$- coordinate to the shock position,
$z\to z_{s}\left(t\right)+z$. The above equation is distinct from
those typically used for particle acceleration in that it includes
an additional cross-field transport with the particle diffusivity
$\kappa_{\perp}.$ This term is needed here because, by contrast to
one-dimensional treatments with the constant injection rate along
the shock surface, the injection intensity $Q\left(r_{\perp}\right)$
vanishes at $r_{\perp}\gtrsim r_{\text{cr}}\left(t\right)$, while
the injection area grows as the shock expands and $r_{\text{cr}}$
increases. Already for this reason, the \emph{acceleration process
is fundamentally time-dependent}. The above equation is written in
the shock frame. Although this frame is not inertial, since $\dot{z}_{s}=u_{shock}\neq const$,
eq.(\ref{eq:CDeq}) describes a mass-less CR-fluid diffusively coupled
to the local flow, so it is valid for $u=u\left(t\right)$. Here $p_{0}$
refers to the particle injection momentum.

\section{Loss-Free Time-Dependent Spectrum of Accelerated Particles\label{sec:Solutions}}

Now we focus on the solution of eq.(\ref{eq:CDeq}) that determines
both the emission spectrum of accelerated particles observed along
the line of sight, as shown in Figs.\ref{fig:Spherical-shock-surface}a,b,
and a CR production spectrum integrated over the active life of an
SNR shock in question. In this paper, we will address the latter task
by calculating a radially integrated spectrum. It is a good proxy
for the LoS-integrated spectrum, but they are not identical except
when the local spectrum is radially independent. The radially integrated
spectrum can be derived from eq.(\ref{eq:CDeq}) directly.

\subsection{Radially integrated spectrum}

As we pointed out in the preceding section, the particle density presumably
vanishes at the critical radius, $r_{\perp}=r_{cr}\left(t\right).$
Therefore, we may extend the integration to infinity, thus introducing
an integrated spectrum and the source of injected particles as

\begin{equation}
\bar{F}\left(z,p,t\right)=2\pi\int_{0}^{\infty}r_{\perp}f\left(z,r_{\perp},p,t\right)dr_{\perp}\;\;\;\;{\rm and}\;\;\;\;\;S\left(t\right)=2\pi\int_{0}^{\infty}r_{\perp}Q\left(r_{\perp},t\right)dr_{\perp},\label{eq:FbarDef}
\end{equation}
With these definitions, from eq.(\ref{eq:CDeq}) we obtain

\begin{equation}
\frac{\partial\bar{F}}{\partial t}+u\frac{\partial\bar{F}}{\partial z}-\frac{\partial}{\partial z}\kappa_{\parallel}\frac{\partial\bar{F}}{\partial z}=\frac{1}{3}\frac{\partial u}{\partial z}p\frac{\partial\bar{F}}{\partial p}+S\left(t\right)\delta\left(z\right)\delta\left(p-p_{0}\right).\label{eq:raveragedDC}
\end{equation}
It is implied here, that $u$ and $\kappa_{\parallel}$ do not depend
on $r_{\perp}$. This assumption is acceptable for $\kappa_{\parallel}$
only in the area of particle localization, $r_{\perp}\le r_{cr}$.
The parallel diffusion strongly increases outside of this area where
the turbulence level is expected to be much lower. Conversely, the
perpendicular transport strongly decreases beyond the point $r_{\perp}=r_{cr}.$
The interface between different propagation regimes near $r_{\perp}=r_{cr}$,
which in general is $p$ and $z$-dependent, deserves a separate study.
Here, we assume that the particle density drops beyond $r_{\perp}=r_{cr}$
sharply, and accept the value $\kappa_{\parallel}\left(r_{\perp}<r_{cr}\right)\approx const$
in eq.(\ref{eq:raveragedDC}), but allow for an arbitrary momentum
dependence, $\kappa_{\parallel}\left(p\right)$. We will also neglect
its strong enhancement that potentially can result in fast particle
losses along the field line in the area $r_{\perp}\gtrsim r_{cr}$.
Based on this assumption, the radially integrated convection-diffusion
equation, eq.(\ref{eq:raveragedDC}) does not contain radial losses
(see, however, Sec.\ref{sec:Comparison-with-Standard}). Its solution
will thus constitute the flattest possible, loss-free spectrum. Yet
as we will see, the spectrum shows an evident rollover towards higher
momenta due to the time-dependent acceleration effects discussed in
Sec.\ref{sec:DSA-Mechanism-With}. The neglected losses would further
steepen the spectrum, which we plan to address in a separate publication.
Next, we discuss what stage of an SNR expansion is most relevant to
our subject.

\subsection{Selection and Description of the SNR Expansion Stage\label{subsec:SNR stage choice}}

From the known stages of SNR dynamical evolution \citep{Chevalier1977ARA&A,Bisnovatyi95,McKeeTruelove95},
the strongest impact on the CR production must have the ejecta-dominated
(ED) and Sedov-Taylor (ST) stages. The shock radius grows linearly,
$R_{s}\propto t$, during the ED stage, and slows down considerably
during the ST stage, $R_{s}\propto t^{2/5}.$ The transition between
the two is smooth, nominally occurring at $t\sim t_{ST}$, which we
will further denote by $t_{0}$, thus only loosely associating it
with the commonly defined quantity, $t_{ST}\approx0.495M_{e}^{5/6}/\rho_{0}^{1/3}\sqrt{E}$$\approx209\left(M_{e}/M_{\odot}\right)^{5/6}/n_{0}^{1/3}\sqrt{E_{51}}$.
Here $M_{e}$ and $M_{\odot}$ are the ejecta and solar masses, respectively,
$E$ is the ejecta energy (given in $10^{51}$erg units $E_{51}$),
$n_{0}=\rho_{0}/2.34\times10^{-24}\text{g}$, with $\rho_{0}$ being
the ambient mass density. \citet{McKeeTruelove95} give a convenient
analytic fit to the actual hydrodynamic solution of an SNR expansion
problem well describing the first two stages of expansion:

\begin{equation}
R_{s}=R_{ST}\left\{ \begin{array}{cc}
1.37t\left(1+0.6t^{3/2}\right)^{-2/3}, & t<1\\
\left(1.56t-0.56\right)^{2/5}, & t>1
\end{array}\right.\label{eq:STcasesED}
\end{equation}
\begin{wrapfigure}{o}{0.5\columnwidth}%
\includegraphics[scale=0.6]{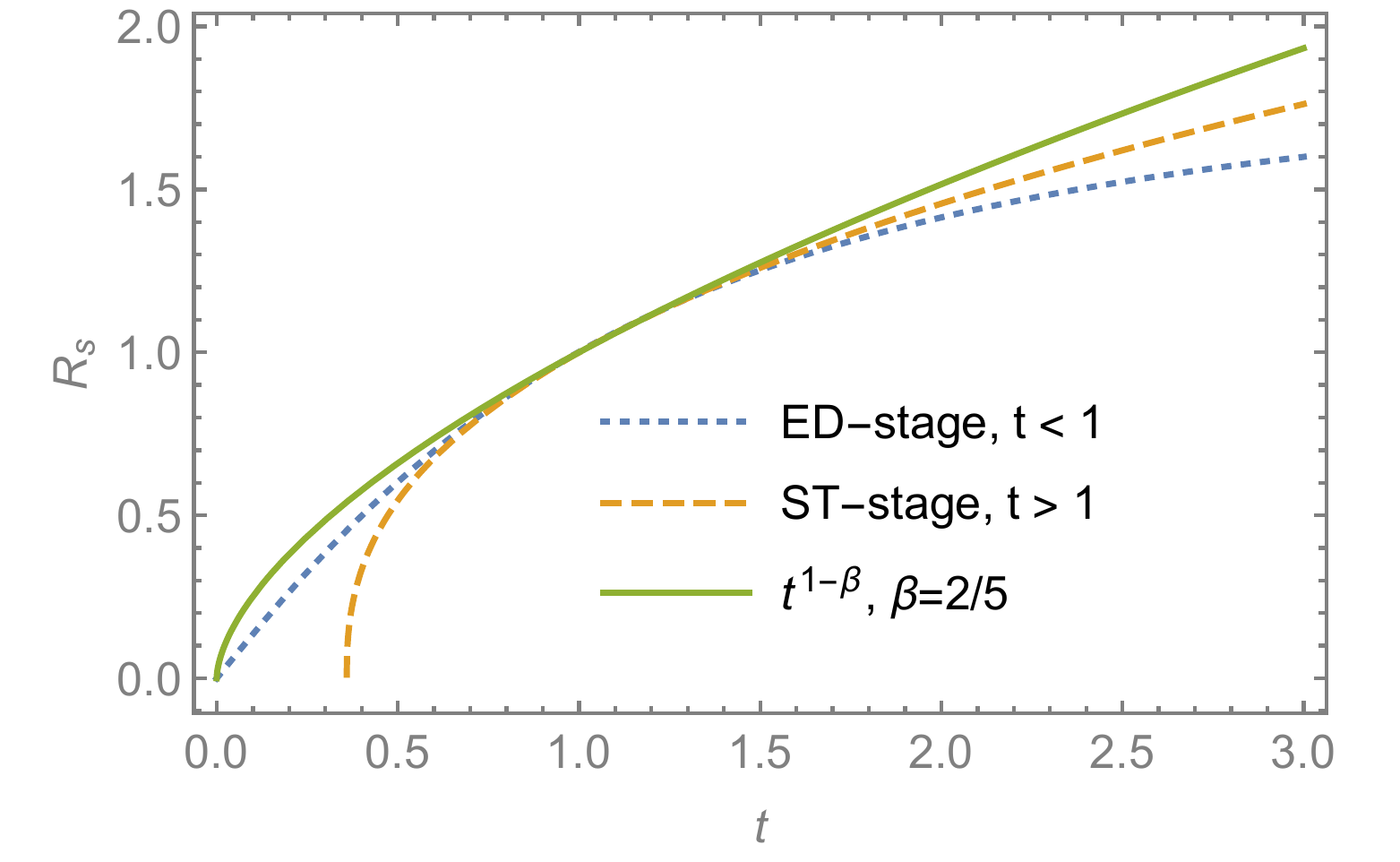}

\caption{Shock radius (normalized to $R_{\text{ST}}$) as a function of time
(in units of $t_{\text{ST}})$ shown using three different approximations,
indicated on the plot.\label{fig:Shock-radius-(normalized}}
\end{wrapfigure}%
where $t$ is given in units of $t_{\text{ST}}$ ($t_{0}$ in our
nomenclature) and $R_{\text{ST}}\approx0.727M_{e}^{1/3}\rho_{0}^{-1/3}\approx2.23\left(M_{e}/M_{\odot}\right)^{1/3}n_{0}$
pc. However, most important for the CR production is perhaps the transient
part between the ED and ST stages. By this epoch, the shock radius
has grown large enough to process a significant amount of interstellar
material while the shock is still strong enough to accelerate particles
efficiently. Just a single power-law, $R_{s}/R_{\text{ST}}=t^{1-\beta}$,
with $\beta\approx2/5,$ reproduces the transition reasonably well,
by growing slower than ED but faster than ST stage. All the three
approximations are shown in Fig.\ref{fig:Shock-radius-(normalized}.
The single power-law approximation substantially deviates from the
more accurate representations in eq.(\ref{eq:STcasesED}) only early
in the free-expansion (ED) stage, $t\ll t_{\text{ST}}$, or later,
$t\gg t_{\text{ST}},$ in the ST stage. These two regimes in the SNR
expansion, however, have a less profound impact on the CR production
for the reasons mentioned earlier, so we can employ the single power-law
approximation. A big advantage of this simplification is that it leads
to an \emph{exact} solution of eq.(\ref{eq:raveragedDC}).

Although we will solve eq.(\ref{eq:raveragedDC}) for arbitrary $\beta$
we recapitulate that $\beta\approx2/5$ value is a good choice based
on observations of the most popular remnants, especially the SN 1006,
as shown in Table \ref{tab:Expansion-indices-and}, also taken from
\citet{McKeeTruelove95}. The actual expansion index, $1-\beta$ is
consistently larger than the ST value of $1-\beta=0.4$ for all of
the listed objects, so we will use the value $1-\beta=3/5$ for the
illustrations of our solution.

\subsection{Solution of Convection-Diffusion Equation. \label{subsec:Solution-of-Convection-Diffusion}}

\subsubsection{Spatial Profile}

For the function $u\left(z\right)$, we make the conventional plane-shock
assumption: $u=-u_{1}$, for $z>0$ and $u=-u_{2}$, for $z\le0$,
where $u_{1}>u_{2}>0$. As the shock velocity decays in time, we write

\[
u=-\left(\frac{t_{0}}{t}\right)^{\beta}\left\{ \begin{array}{cc}
u_{1}, & z>0\\
u_{2}, & z\le0
\end{array}\right.
\]
In the Sedov-Taylor stage, for example, $\beta=3/5,$ so that the
shock propagates at the speed $U_{s}\propto t^{-3/5}$ and its radius
grows as $R_{s}\propto t^{2/5}.$

\begin{wraptable}{o}{0.5\columnwidth}%
\begin{tabular}{|c|r|r|}
\hline 
SNR & $t/t_{\text{ST}}$ & $1-\beta$\tabularnewline
\hline 
\hline 
Cas A & 0.57-2.3 & 0.79-0.48\tabularnewline
\hline 
Kepler & 0.44 & 0.85\tabularnewline
\hline 
Tycho & 0.83 & 0.69\tabularnewline
\hline 
SN 1006 & 1.4 & 0.54\tabularnewline
\hline 
\end{tabular}

\caption{Expansion indices and ages of selected SNRs. The data are taken from
the review by \citet{McKeeTruelove95}, where the references to the
sources are provided.\label{tab:Expansion-indices-and}}
\end{wraptable}%

The choice of $\kappa_{\parallel}\left(p,t\right)$ is more difficult.
Ideally, its functional form should be obtained self-consistently
with the spectrum of accelerated particles, since the particle-driven
turbulence determines their diffusivity. Not knowing the turbulence
beforehand, we still can guess $\kappa_{\parallel}$ on dimensional
considerations. Given the shock speed $U_{s}$ and its radius $R_{s}$,
the only combination of the two with $\kappa_{\parallel}$-dimensionality
is $\kappa_{\parallel}\sim R_{s}U_{s}\propto t^{1-2\beta}$. If we
substitute $U_{s}\sim10^{3}km/s$ and $R_{s}\sim1pc$, then $\kappa_{\parallel}\sim10^{26}cm^{2}/s$,
which may be tentatively accepted as a turbulently suppressed CR diffusivity
in the ISM by some two orders of magnitude. However, there are some
caveats here. Not only we do not know the coefficient for $\kappa_{\parallel}$,
but its possibly strong dependence on $p$ is unknown. The clue comes
from an obvious constraint on the highest energy particles with momentum
$p_{{\rm max}}$. They cannot diffuse in course of acceleration farther
than say $\kappa_{\parallel}\left(p_{max}\right)/U_{s}<0.1R_{s}$.
This constraint fixes the uncertain factor at $0.1$, somewhat arbitrarily
but not unreasonably and, in addition, it provides some implications
for $\kappa_{\parallel}\left(p\right)$ dependence. An equivalent
result can be obtained from the requirement that the acceleration
time to $p_{max}$, $\tau_{a}\sim\kappa_{\parallel}\left(p_{max}\right)/U_{s}^{2}$
is shorter than the shock dynamical time, $R_{s}/U_{s}$ \citep{DruryEscape11}.
That being said, we do not need to specify the magnitude of $\kappa_{\parallel}\left(p\right)$
as we will obtain the solution keeping it arbitrary. All we need is,
by adopting the scaling $\kappa_{\parallel}\propto U_{s}R_{s}$, to
fix the time-dependent part of the particle diffusion coefficient:

\[
\kappa_{\parallel}\left(p,t\right)=\kappa\left(p\right)\left(\frac{t}{t_{0}}\right)^{1-2\beta}
\]
Again, it is difficult to determine the $\kappa\left(p\right)$ dependence
as it is linked to the spectrum of turbulence, unknown until the particle
spectrum is obtained from eq.(\ref{eq:raveragedDC}). Fortunately,
as we find the DSA solution for an arbitrary $\kappa\left(p\right)$,
the two complicated problems will at least be separated.

\begin{figure}
\includegraphics[scale=0.55]{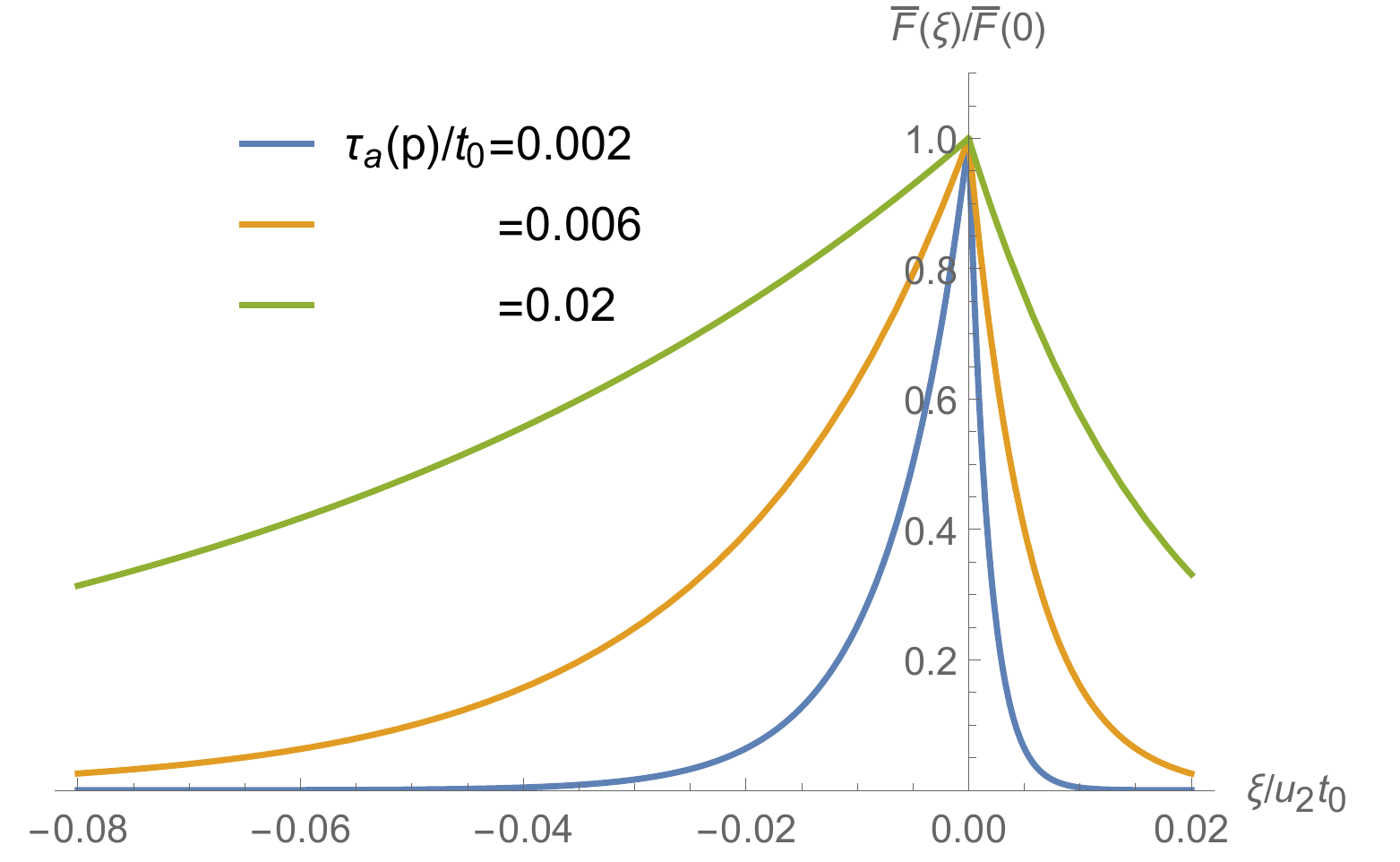}\includegraphics[scale=0.55]{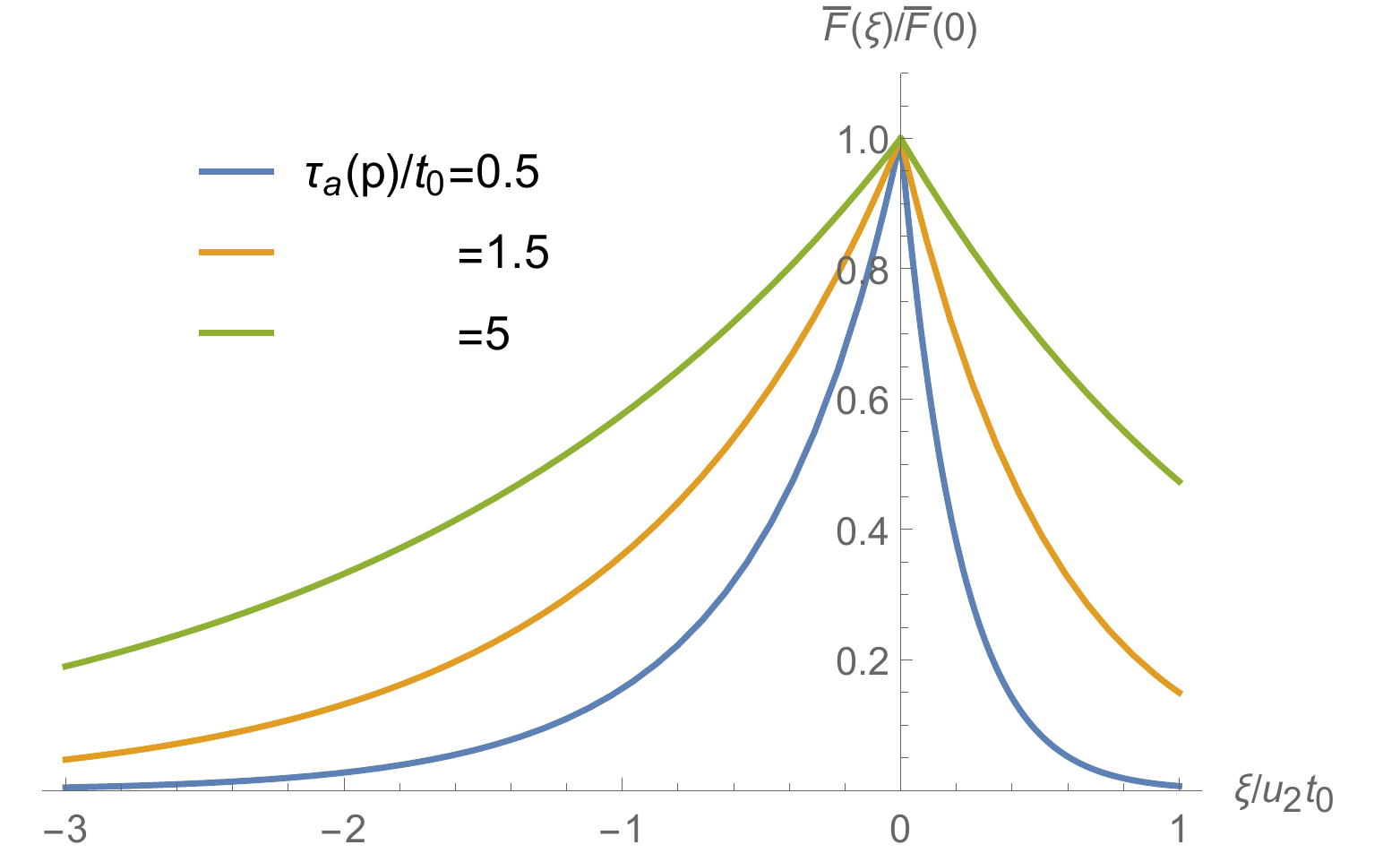}\caption{Spatial profiles of upstream and downstream particle distributions
$\bar{F}_{u,d}$ normalized to $\bar{F}_{0}\left(p\right)$ (shock
value), depending on the dimensionless distance from the shock front
$\xi/u_{2}t_{0}=z/u_{2}t_{0}^{2/5}t^{3/5}$ ($\xi>0$- upstream, $\xi\le0$-
downstream). The left/right panel shows the profiles in the low/high
-momentum range presented in the form of acceleration time to the
respective momentum $\tau_{a}=2\left(1-\beta\right)\kappa\left(p\right)/u_{2}^{2}$.\label{fig:Spatial-profiles-of}}
\end{figure}

\begin{figure}
\includegraphics[scale=0.55]{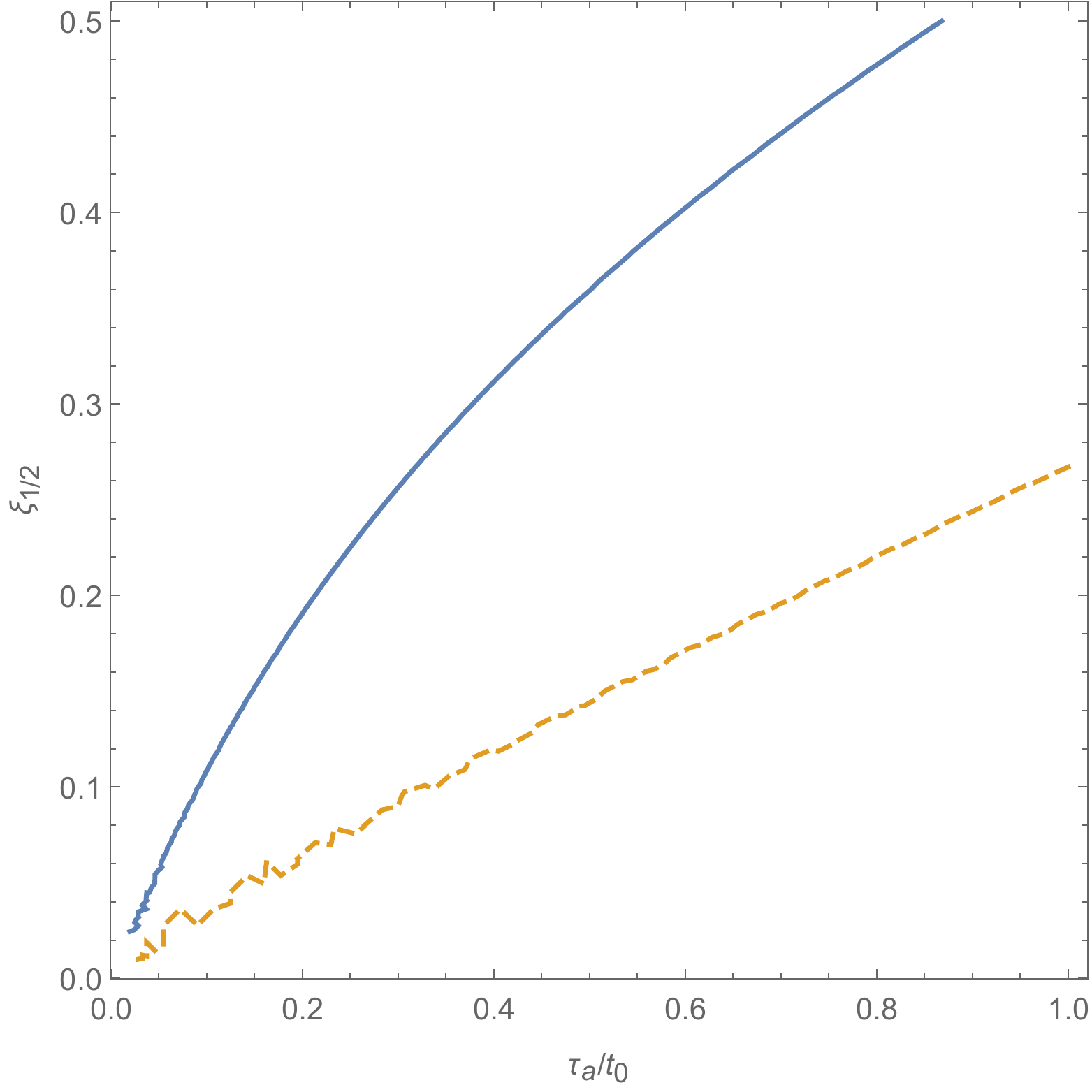}\includegraphics[scale=0.54]{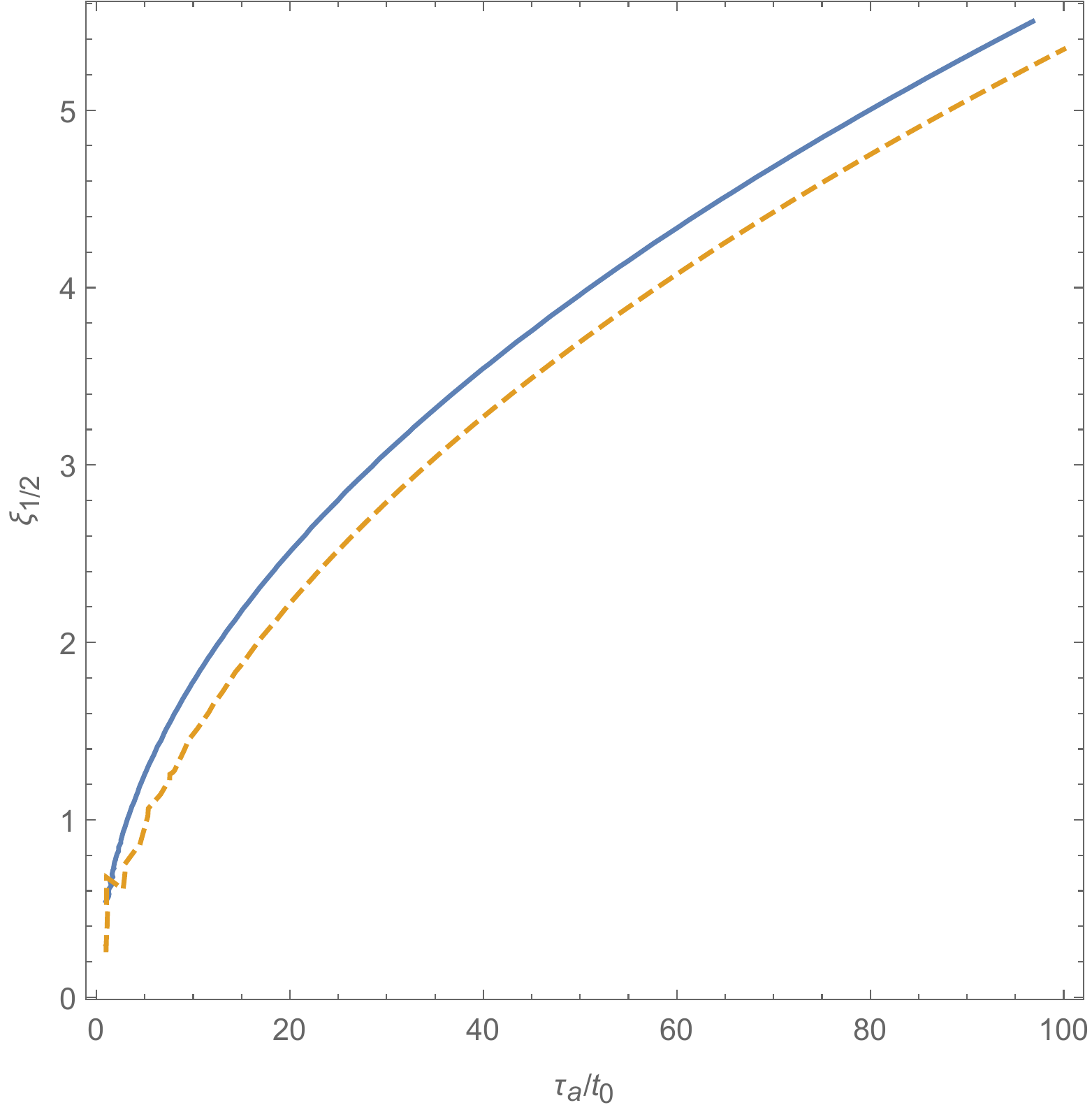}

\caption{Half-width of particle distribution ($\xi_{1/2}$- distance from the
shock where $\bar{F}$ decreases by half from its maximum at $\xi=0$)
upstream (dashed lines) and downstream (solid lines) depending on
momentum, expressed in the acceleration time $\tau_{a}=2\left(1-\beta\right)\kappa\left(p\right)/u_{2}^{2}$.\label{fig:Half-width-of-particle}}
\end{figure}

We seek a self similar solution to eq.(\ref{eq:raveragedDC}) in the
following form

\[
\bar{F}\left(t,z,p\right)=\phi\left(t,p\right)F\left(p,\xi\right)
\]
where

\[
\xi=z\left(\frac{t}{t_{0}}\right)^{\beta-1}.
\]
For $z,\xi\neq0$ eq.(\ref{eq:raveragedDC}) then rewrites

\begin{equation}
\kappa\frac{d^{2}F}{d\xi^{2}}+\left(u_{1,2}+a\xi\right)\frac{\partial F}{\partial\xi}-\zeta F=0\label{eq:DCeqForFofKsi}
\end{equation}
where $u_{1,2}$ values refer to the upstream ($\xi\ge0$) and downstream
($\xi<0$) half-space, respectively. We introduced the following two
parameters

\begin{equation}
\zeta=\frac{t}{t_{0}\phi}\frac{\partial\phi}{\partial t},\;\;\;\;\;\;a=\frac{1-\beta}{t_{0}}\label{eq:zetaANDaDEF}
\end{equation}
We will also use the above equation in the following form suitable
for all $\xi$

\begin{equation}
\kappa\frac{d^{2}F}{d\xi^{2}}+\left[U\left(\xi\right)+a\xi\right]\frac{\partial F}{\partial\xi}-\zeta F=\frac{1}{3\phi}\frac{\partial U}{\partial\xi}p\frac{\partial}{\partial p}\left(\phi F\right)-\left(\frac{t}{t_{0}}\right)^{\beta}\frac{S\left(t\right)}{\phi}\delta\left(\xi\right)\delta\left(p-p_{0}\right)\label{eq:FofKsiEq}
\end{equation}
where

\[
U\left(\xi\right)=u_{2}+\left(u_{1}-u_{2}\right)H\left(\xi\right)
\]
and $H\left(\xi\right)$ is the Heaviside unit function ($H^{\prime}\left(x\right)=\delta\left(x\right)$).
The solution of eq.(\ref{eq:DCeqForFofKsi}) can be expressed in parabolic
cylinder functions. To select the solutions with correct asymptotic
properties upstream and downstream, it is convenient to further transform
it using the following substitutions

\begin{equation}
F=g\left(\xi,p\right)\exp\left[-\frac{\xi}{2\kappa}\left(u_{1,2}+a\frac{\xi}{2}\right)\right]=g\left(\eta\right)\exp\left[-\frac{\eta^{2}}{4}+\frac{u_{1,2}^{2}}{4\kappa a}\right]\label{eq:FtogTransf}
\end{equation}

\begin{equation}
\xi=\sqrt{\kappa/a}\eta-u_{1,2}/a\label{eq:ksiOfetaDef}
\end{equation}
which bring us to the canonical form of eq.(\ref{eq:DCeqForFofKsi})

\begin{equation}
\frac{\partial^{2}g}{\partial\eta^{2}}-\left(\frac{1}{4}\eta^{2}+\frac{\zeta}{a}+\frac{1}{2}\right)g=0\label{eq:gOfetaEq}
\end{equation}
Note that the range of the new variable $\eta$ is $u_{1}/\sqrt{\kappa a}<\eta<\infty$,
upstream and $-\infty<\eta<u_{2}/\sqrt{\kappa a}$ downstream. We
observe that the shock coordinate $\xi=0$ maps to different endpoints
$\eta$ in the above intervals, because of the discontinuity in $U\left(\xi\right)$
at $\xi=0$. By denoting $W=\eta^{2}/4+\zeta/a+1/2$, we can write
the asymptotic expressions at $\eta\to\pm\infty$ for the two linearly
independent solutions of eq.(\ref{eq:gOfetaEq}) as follows

\[
g_{\pm}\propto\frac{1}{W^{1/4}}\exp\left(\pm\int\sqrt{W}d\eta\right)
\]
Because $F\left(p,\xi\right)$ must vanish at $\left|\xi\right|\to\infty$,
we discard one of the solutions $g_{+}$ or $g_{-}$, depending on
what branch of $\sqrt{W}$ is chosen. Thus, we must select for $g\left(\eta\right)$
the parabolic cylinder function $D_{\nu}\left(\eta\right)$ for $\eta>0$
(upstream) with the index $\nu=-\zeta/a-1$ \citep{Bateman1955}.
So, the solution of eq.(\ref{eq:DCeqForFofKsi}) upstream ($\eta\ge u_{1}/\sqrt{\kappa a}$
) must be chosen in the form

\[
F=F_{u}=2D_{-\zeta/a-1}\left(\eta\right)\exp\left(-\frac{1}{4}\eta^{2}\right)
\]
Before writing the respective solution for the downstream medium we
constrain $\zeta/a$ parameter which allows us to simplify the solution.
Indeed, according to eq.(\ref{eq:zetaANDaDEF}), we must constrain
$\phi$ by the relation $t\partial\phi/\partial t\propto\phi,$ while
$S$ must scale with time as

\begin{equation}
S\left(t\right)\propto r_{cr}^{2}\left(t\right)Q\left(t\right)\propto R_{s}^{2}\left(t\right)U_{s}\left(t\right)\propto t^{2-3\beta}\label{eq:Sscaling}
\end{equation}
Comparing this with the last term on the rhs of eq.(\ref{eq:FofKsiEq}),
we obtain the time dependence of $\phi$ as $\phi\propto t^{2\left(1-\beta\right)}.$
Therefore, parameter $\zeta/a$ is $\zeta/a=t\dot{\phi}/\left(1-\beta\right)\phi=2$.
For $\zeta/a=2$ the parabolic cylinder function and the solution
upstream simplify as follows

\begin{equation}
\bar{F}_{u}=2\phi\left(t,p\right)D_{-3}\left(\eta\right)\exp\left(-\frac{1}{4}\eta^{2}\right)=\phi\left[\sqrt{\frac{\pi}{2}}\left(1+\eta^{2}\right){\rm erfc}\left(\frac{\eta}{\sqrt{2}}\right)-\eta\exp\left(-\frac{1}{2}\eta^{2}\right)\right]\label{eq:UpstreamFbar}
\end{equation}
where ${\rm erfc}\left(x\right)=1-{\rm erf}\left(x\right)=\left(2/\sqrt{\pi}\right)\int_{x}^{\infty}\exp\left(-t^{2}\right)dt$
- the complementary error function. The momentum spectrum at the shock
takes the form:
\begin{equation}
\bar{F}_{0}=\phi\left(t,p\right)\left[\sqrt{\frac{\pi}{2}}\left(1+\frac{u_{1}^{2}}{\kappa a}\right){\rm erfc}\left(\frac{u_{1}}{\sqrt{2\kappa a}}\right)-\frac{u_{1}}{\sqrt{\kappa a}}\exp\left(-\frac{u_{1}^{2}}{2\kappa a}\right)\right]\label{eq:FzeroBarThroughfi}
\end{equation}
On the downstream side of the shock we select the second linearly
independent solution of eq.(\ref{eq:gOfetaEq}), namely $D_{-3}\left(-\eta\right)$,
so that the solution $\bar{F}$ downstream that matches the above
expression at the shock takes the following form:

\[
\bar{F}_{d}=2\phi\frac{D_{-3}\left(u_{1}/\sqrt{\kappa a}\right)}{D_{-3}\left(-u_{2}/\sqrt{\kappa a}\right)}D_{-3}\left(-\eta\right)\exp\left(-\frac{1}{4}\eta^{2}+\frac{u_{2}^{2}-u_{1}^{2}}{4\kappa a}\right)
\]
which we rewrite as 
\begin{equation}
\bar{F}_{d}=\phi\left(t,p\right)\Lambda\left(p\right)\left[\sqrt{\frac{\pi}{2}}\left(1+\eta^{2}\right){\rm erfc}\left(-\frac{\eta}{\sqrt{2}}\right)+\eta\exp\left(-\frac{1}{2}\eta^{2}\right)\right]\label{eq:DownstreamFbar}
\end{equation}
with

\[
\Lambda\left(p\right)\equiv\frac{\sqrt{\frac{\pi}{2}}\left(1+u_{1}^{2}/\kappa a\right){\rm erfc}\left(u_{1}/\sqrt{2\kappa a}\right)-\left(u_{1}/\sqrt{\kappa a}\right)\exp\left(-u_{1}^{2}/2\kappa a\right)}{\sqrt{\frac{\pi}{2}}\left(1+u_{2}^{2}/\kappa a\right){\rm erfc}\left(-u_{2}/\sqrt{2\kappa a}\right)+\left(u_{2}/\sqrt{\kappa a}\right)\exp\left(-u_{2}^{2}/2\kappa a\right)}
\]
The spatial profiles of the upstream and downstream solutions $\bar{F}_{u}$
and $\bar{F}_{d}$ given by eqs.(\ref{eq:UpstreamFbar}) and (\ref{eq:DownstreamFbar})
are shown in Fig.(\ref{fig:Spatial-profiles-of}) for different particle
momenta, expressed in a form of acceleration time $\tau_{a}$:

\begin{equation}
\tau_{a}=2\left(1-\beta\right)\kappa\left(p\right)/u_{2}^{2}\label{eq:TauaDef}
\end{equation}
These profiles present a notable contrast to the standard DSA solution
for a steadily propagating shock which has a flat particle distribution
downstream. Although a trend to a broader downstream distribution
is seen in the above solution at lower momenta (more exactly, lower
$\tau_{a}\left(p\right)$), the difference between the upstream and
downstream spatial extent of the particle profile diminishes at higher
momenta. We emphasize this observation using Fig.\ref{fig:Half-width-of-particle}
that shows the half-width $\xi_{1/2}$ (the distance from the front
where $\bar{F}\left(\xi\right)$ drops by 50\%) for the upstream and
downstream distributions, depending on momentum.


\subsubsection{Momentum Spectrum}

Turning to the momentum distribution at the shock front, we integrate
eq.(\ref{eq:FofKsiEq}) across its velocity jump and obtain the following
equation for $\bar{F}_{0}\left(p\right)$:

\begin{equation}
\frac{\Delta u}{3}p\frac{\partial\bar{F}_{0}}{\partial p}-\left.\kappa\frac{\partial\bar{F}}{\partial\xi}\right|_{\xi=0-}^{\xi=0+}=S\left(t\right)\left(\frac{t}{t_{0}}\right)^{\beta}\delta\left(p-p_{0}\right)\label{eq:FzeroBarWithKsi}
\end{equation}
where $\Delta u=u_{1}-u_{2}$. Introducing the following notation

\[
\Phi\left(v\right)=\int_{v}^{\infty}\exp\left[v^{2}-x^{2}\right]dx,\qquad\Psi\left(v\right)=\frac{1-2v\Phi\left(v\right)}{\left(2v^{2}+1\right)\Phi\left(v\right)-v},\,\,\,\,\,\,\,\,\,\,v_{1,2}\left(p\right)\equiv\frac{u_{1,2}}{\sqrt{2\kappa\left(p\right)a}}
\]
eq.(\ref{eq:FzeroBarWithKsi}) can be rewritten as follows

\begin{equation}
-\frac{v_{1}-v_{2}}{3\bar{F}_{0}}p\frac{\partial\bar{F}_{0}}{\partial p}=\Psi\left(v_{1}\right)+\Psi\left(-v_{2}\right)-\frac{S\left(t\right)}{\sqrt{2a\kappa}\bar{F}_{0}}\left(\frac{t}{t_{0}}\right)^{\beta}\delta\left(p-p_{0}\right)\label{eq:SpIndexWithTime}
\end{equation}
 This relation readily provides an exact closed-form expression for
the power-law index of the distribution at the shock front for $p\ge p_{0}$:

\begin{equation}
q\left(p\right)\equiv-\frac{p}{\bar{F}_{0}}\frac{\partial\bar{F}_{0}}{\partial p}=\frac{3\sigma}{\text{\ensuremath{\sigma-1}}}\frac{1}{v_{1}}\left[\Psi\left(v_{1}\right)+\Psi\left(-v_{2}\right)\right]\label{eq:qOfpFinalGen}
\end{equation}
where $\sigma=u_{1}/u_{2}=v_{1}/v_{2}$ and 
\begin{equation}
v_{2}=\frac{u_{2}}{\sqrt{2a\kappa\left(p\right)}}=\sqrt{\frac{t_{0}}{\tau_{a}\left(p\right)}}\label{eq:vOfP}
\end{equation}
The quantity $v_{2}\left(p\right)$ is thus the ratio of the SN dynamic
timescale $t_{0}$ to the characteristic time of acceleration to the
momentum $p$: $\tau_{a}=2\left(1-\beta\right)\kappa\left(p\right)/u_{2}^{2}$.
The time variable formally enters the last term on the r.h.s. of eq.(\ref{eq:SpIndexWithTime}).
However, we established earlier, eq.(\ref{eq:Sscaling}), that $S\propto t^{2-3\beta}$
and $\bar{F}_{0}\propto\phi\propto t^{2\left(1-\beta\right)}$, so
$t$ cancels out from eq.(\ref{eq:SpIndexWithTime}). By writing then
$S=S_{0}\left(t/t_{0}\right)^{2-3\beta},$ and using eqs.(\ref{eq:SpIndexWithTime})
and (\ref{eq:qOfpFinalGen}) we obtain the spectrum of accelerated
particles at the shock front:

\begin{equation}
\bar{F}_{0}=\frac{3S_{0}}{\Delta up_{0}}\left(\frac{t}{t_{0}}\right)^{2\left(1-\beta\right)}\exp\left[-\int_{p_{0}}^{p}q\left(p\right)\frac{dp}{p}\right]\label{eq:F0barGeneral}
\end{equation}
Recall that $\bar{F}$ is a radially integrated spectrum, according
to eq.(\ref{eq:FbarDef}), so the factor $t^{2\left(\beta-1\right)}$
is the acceleration area, since $R_{SNR}\propto t^{1-\beta}.$ Having
this at hand, the spatial distributions upstream and downstream can
be obtained using eqs.(\ref{eq:UpstreamFbar}-\ref{eq:DownstreamFbar}).
Next, we discuss the momentum spectrum and consider its two simple
limits.

\section{Spectral properties in limiting cases\label{sec:Spectral-properties-and}}

\begin{wrapfigure}{o}{0.5\columnwidth}%
\includegraphics[scale=0.47]{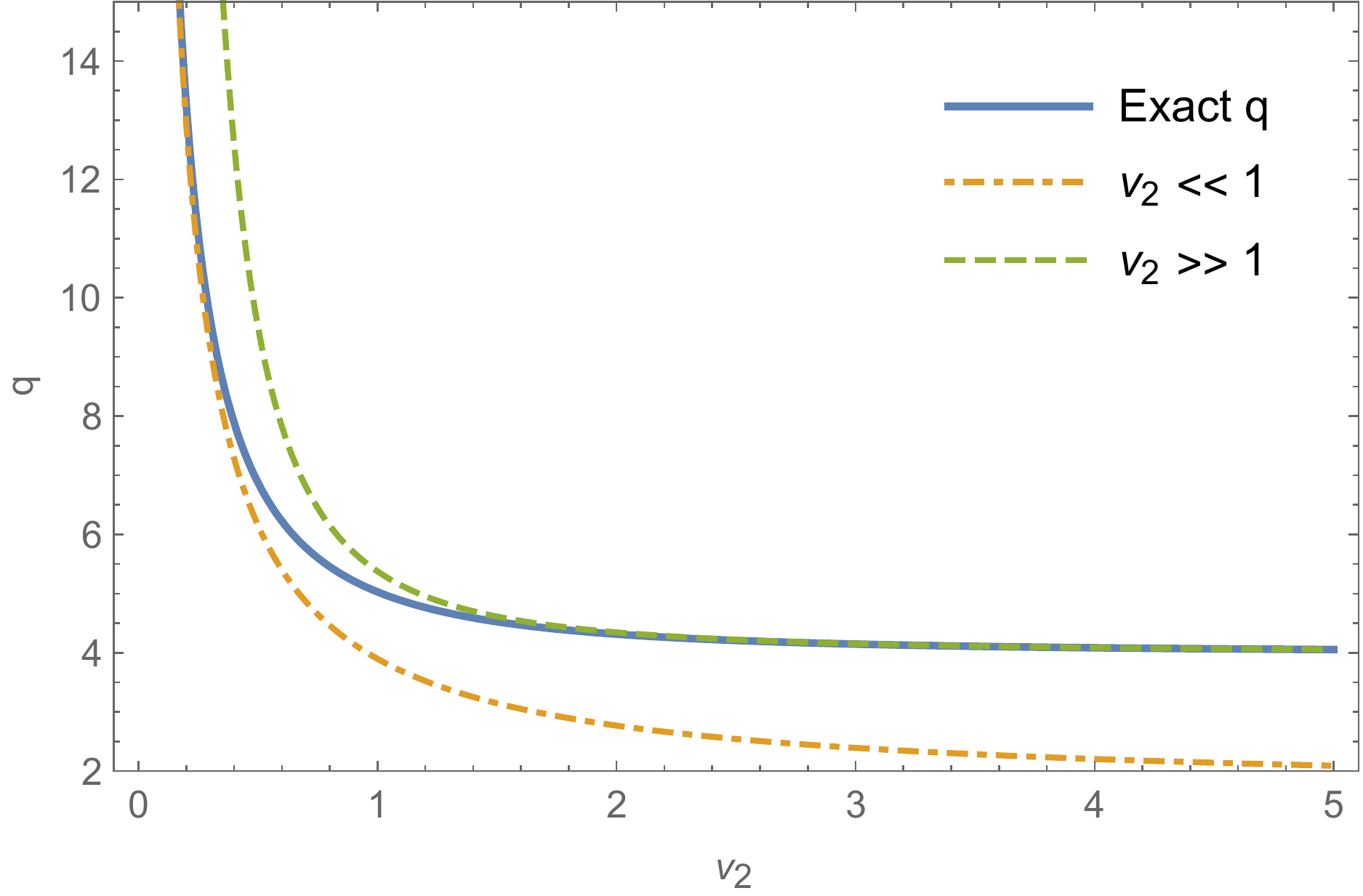}

\caption{Spectral index $q$ of accelerated particles at the shock front as
a function of $v_{2}=\sqrt{t_{0}/\tau_{a}\left(p\right),}$ as given
by eqs.(\ref{eq:qOfpFinalGen}) and (\ref{eq:vOfP}). Also shown by
the dashed and dashed-dotted curves the low- and high-momentum approximations
of eq.(\ref{eq:qOfpFinalGen}).\label{fig:Spectral-index-Of-v}}
\end{wrapfigure}

\begin{wrapfigure}{o}{0.5\columnwidth}%
\includegraphics[scale=0.47]{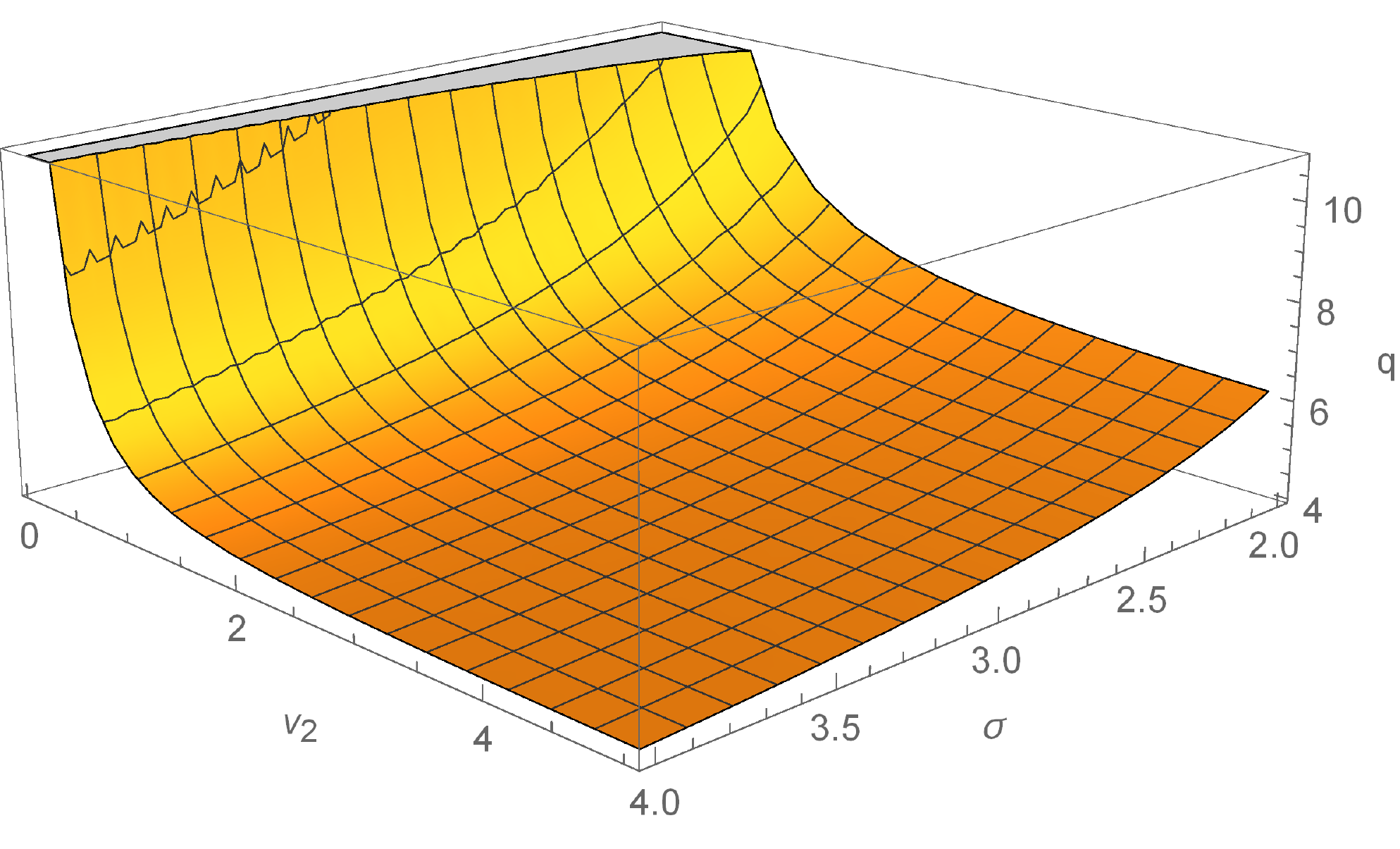}\caption{Surface plot of the spectral index $q\left(\sigma,v_{2}\right)$ shown
as in eq.(\ref{eq:qOfpFinalGen}).\label{fig:Surface-plot-of-q}}
\end{wrapfigure}

The spectral index of accelerated particles in eq.(\ref{eq:qOfpFinalGen})
depends on the particle momentum through the parameter $v_{2}\left(p\right)\equiv\sqrt{t_{0}/\tau_{a}\left(p\right)}\propto\kappa^{-1/2}\left(p\right)$,
which is not surprising. For the time-dependent acceleration, the
dynamical timescale of the accelerator, $t_{0},$ enters the result
in the form of its ratio to the acceleration time. A stationary loss-free
accelerator, on the contrary, does not have any timescale, to be compared
with the particle acceleration time, $\tau_{a}\left(p\right)$. Hence
its spectral index is \emph{momentum-independent}. However, in the
limit $\tau_{a}\ll t_{0}$ there should be no difference between the
two cases. Indeed, we can simplify the general solution for $q\left(p\right)$
in eq.(\ref{eq:qOfpFinalGen}) for this case and for the case of long
acceleration time, $\tau_{a}\gg t_{0}$ suitable, respectively, for
low and high particle momenta. The last statement implies that the
diffusion coefficient $\kappa$ grows with $p$, but in general, we
have imposed no restrictions on $\kappa\left(p\right)$. For $v\gg1$
(low $p$) we have

\begin{equation}
q\approx\frac{3\sigma}{\sigma-1}\left(1+\frac{3+2\sigma}{2\sigma^{2}v_{2}^{2}}\right)\label{eq:LowEnSp}
\end{equation}
which converges to the conventional strong shock result $q=3\sigma/\left(\sigma-1\right)$
when $v_{2}\to\infty$, as expected. In the opposite case of $v_{2}\ll1$,
one obtains

\begin{equation}
q=\frac{12}{\sqrt{\pi}\left(\sigma-1\right)}\frac{1}{v_{2}}+6\left(\frac{4}{\pi}-1\right)\label{eq:qAsymLargep}
\end{equation}
Note that the asymptotically dominant term is the first one. It is
responsible for an exponential decay of the spectrum at high momenta

\begin{equation}
F\propto p^{-q_{0}}e^{-q_{1}\left(p\right)}\label{eq:SpectrumLargeP}
\end{equation}
where $q_{0}=6\left(4/\pi-1\right)$. Assuming the Bohm regime, $v_{2}=\sqrt{t_{0}/\tau_{a}}\propto1/\sqrt{p}$,
for $q_{1}$ we have $q_{1}=\left[24/\sqrt{\pi}\left(\sigma-1\right)\right]\sqrt{\tau_{a}/t_{0}}\propto\sqrt{p}.$
Note that the above spectrum decays slower at high momenta than the
standard DSA would for the same $\kappa\propto p$ diffusion scaling.
We will compare the two spectra in the next section. The spectral
indices given by eqs.(\ref{eq:LowEnSp}) and (\ref{eq:qAsymLargep})
are shown in Fig.\ref{fig:Spectral-index-Of-v} in a strong shock
limit, $\sigma=4$, along with the exact result, given in eq.(\ref{eq:qOfpFinalGen}).
The index $q\left(\sigma,v_{2}\right)$ for arbitrary $\sigma$ and
$v_{2}$ is shown in Fig.\ref{fig:Surface-plot-of-q} as a surface
plot.

From these results, we conclude that even neglecting particle losses
one obtains a significant spectral steepening. It is accounted for
by broadening the particle injection area in time, as discussed in
Sec.\ref{sec:DSA-Mechanism-With}, and slowing down the pace of acceleration.
A more realistic treatment including the boundary losses will result
in a steeper spectrum. That being said, the exponential spectral decline
obtained above is significantly different from the usual DSA result.
We will demonstrate this in the next section.

\section{Comparison with Standard DSA\label{sec:Comparison-with-Standard}}

The purpose of this section is to understand whether the above modification
to the DSA spectrum is a step toward reconciliation with observations.
Because the ``standard'' DSA does not include effects addressed
in this paper, it is necessary to delineate the ``common territory''.
According to the DSA, the spectrum terminates with a smeared cut-off,
rather than a more gradual steepening described in the previous two
sections. We are now poised to contrast these two spectral steepening
mechanisms.

The ordinary DSA cut-off results from either a limited acceleration
time or particle escape due to a finite spatial extent of acceleration
region. In the first case, one can write the following equation for
the maximum momentum

\[
\frac{dp_{\max}}{dt}=\frac{p_{\max}}{t_{acc}}\sim\frac{p_{\max}}{\kappa\left(p_{\max}\right)}U_{sh}^{2}\left(t\right)
\]
This equation can be integrated straightforwardly, even for an arbitrary
$t_{acc}\left(p\right)$, but it does not provide the form of the
spectrum near the cut-off, $p\sim p_{\max}$, which we need to compare
with our results. \citet{Axford1981_ICRC}, for example, imposes an
abrupt cut-off, $f\left(p,t\right)\propto p^{-q_{{\rm st}}}H\left(p_{\max}\left(t\right)-p\right)$
on the DSA solution with $q_{{\rm st}}=3\sigma/\left(\sigma-1\right)$.
Calculations that capture the spectral shape in the cutoff area \citep{Toptygin80,PtusPrish1981}
are limited to momentum independent particle diffusivity $\kappa$
and impose other restrictions on $\kappa$ or the SNR expansion rate
$1-\beta$. In the case of momentum-independent $\kappa$ our self-similar
solution does not steepen with $p$, as it is subjected to the changing
acceleration conditions rather than its finite duration. Including
the latter effect would require a numerical solution of eq.(\ref{eq:raveragedDC})
that should merely demonstrate the convergence to the self-similar
solution with a time-dependent cut-off.

The second approach to the DSA termination \citep{BlandEich87} is
based on the particle escape, which is both better suited to our model,
due to its spatial finiteness, and easier to handle. Indeed, in transforming
from eq.(\ref{eq:CDeq}) to eq.(\ref{eq:raveragedDC}) we have omitted
the particle escape term that we now include as follows

\begin{equation}
\Lambda\bar{F}+u\frac{\partial\bar{F}}{\partial z}-\frac{\partial}{\partial z}\kappa_{\parallel}\frac{\partial\bar{F}}{\partial z}=\frac{1}{3}\frac{\partial u}{\partial z}p\frac{\partial\bar{F}}{\partial p}\label{eq:FbarWithLossTerm}
\end{equation}
Here we have replaced the diffusive flux through $r_{\perp}=r_{{\rm cr}}$
by %
\begin{equation}
-\left.r\kappa_{\perp}\frac{\partial f}{\partial r}\right|_{r=r_{{\rm cr}}}=\Lambda\bar{F}\label{eq:LambdaDef}
\end{equation}
and we will use $\Lambda\sim\kappa_{\perp}/r_{{\rm cr}}^{2}$ for
an estimate. The above formula is a direct analog of the Newton's
law for a flux (e.g., thermal) at an interface between two media (here
at $r=r_{{\rm cr}}$). We have neglected the time derivative, as the
spectrum is expected to be nearly stationary near the cut-off momentum,
since the acceleration should be balanced by particle losses. For
the same reason, the flow velocity and particle diffusivity can also
be considered fixed in time at their current values. The particle
injection term is irrelevant at these energies and therefore omitted.
Solving the above equation upstream and downstream with $u=-u_{1,2},$
and integrating it across the velocity jump, one obtains the following
expression for the spectral slope at the shock
\begin{equation}
\frac{\partial\ln\bar{F}_{0}}{\partial\ln p}=-\frac{3}{2\Delta u}\left[u_{1}\left(1+\sqrt{1+\frac{4\Lambda\kappa_{\parallel}}{u_{1}^{2}}}\right)+u_{2}\left(-1+\sqrt{1+\frac{4\Lambda\kappa_{\parallel}}{u_{2}^{2}}}\right)\right]\label{eq:qWithLossesGeneral}
\end{equation}
Recalling that $\Lambda\sim\kappa_{\perp}/r_{{\rm cr}}^{2}$ we see
that for the low-energy end the parameters $\Lambda\kappa_{\parallel}/u_{1,2}^{2}\sim\kappa_{\perp}\kappa_{\parallel}/r_{{\rm cr}}^{2}u_{1,2}^{2}\ll1$,
which is similar to the limit $v_{2}\gg1$ in eq.(\ref{eq:LowEnSp}).
Note, that we can identify $r_{{\rm cr}}$ in the definition of $\Lambda$
as $r_{{\rm cr}}\approx u_{1}t_{0}/\sqrt{2}.$ In this limit the last
results simplifies to the following

\begin{equation}
\frac{\partial\ln\bar{F}_{0}}{\partial\ln p}\approx-\frac{3u_{1}}{\Delta u}-3\frac{\kappa_{\parallel}\Lambda}{\Delta u}\left(\frac{1}{u_{1}}+\frac{1}{u_{2}}\right)\label{eq:qDSAwithLossesLowMom}
\end{equation}
The first term on the rhs represents the usual DSA slope, coinciding
with the respective result in eq.(\ref{eq:LowEnSp}) valid for moderate
momenta where $\kappa\left(p\right)\ll u^{2}t_{0}$. There is an agreement
in the leading ``standard DSA'' term $3\sigma/\left(\sigma-1\right)$
between eq.(\ref{eq:LowEnSp}) and (\ref{eq:qDSAwithLossesLowMom}).
However, the spectrum steepens significantly slower with $p$ for
the loss-dominated solution in eq.(\ref{eq:qDSAwithLossesLowMom}).
Indeed, the steepening correction to the standard slope is $\Delta q_{L}\sim\kappa_{\parallel}\kappa_{\perp}/r_{{\rm cr}}^{2}u^{2}\propto\kappa_{B}^{2}\propto p^{2}$,
whereas the index steepening due to the time-dependent effects, according
to eq.( \ref{eq:LowEnSp}) is $\Delta q_{TD}$$\sim\kappa_{\parallel}/r_{{\rm cr}}u\sim$$\sqrt{\Delta q_{L}}\gg\Delta q_{L}$,
which grows as $p$, assuming the Bohm diffusivity for $\kappa_{\parallel}$,
but it may grow slower or faster than that depending on $\kappa_{\parallel}\left(p\right)$.

Turning to higher $p$, from eq.(\ref{eq:qWithLossesGeneral}) we
obtain
\[
\frac{\partial\ln\bar{F}_{0}}{\partial\ln p}\approx-\frac{6}{\Delta u}\left(\sqrt{\Lambda\kappa_{\parallel}}\right)-\frac{3}{2}
\]
which grows as $p$, whereas the respective result for the time-dependent
self-similar solution shows only a $p^{1/2}$ growth with momentum
(assuming Bohm regime in both cases). It follows then that the spectrum
steepening with growing momentum occurs more gradually in the case
of a self-similar than steady-state solution with losses. It starts
at lower momenta and continues to higher momenta without sharp termination,
characteristic of the loss-dominated case.

To make a quantitative comparison, let us rewrite the expressions
under the square roots in eq.(\ref{eq:qWithLossesGeneral}) as $4\Lambda\kappa_{\parallel}=\rho p^{2}$,
as to a good approximation $\kappa_{\parallel}\kappa_{\perp}\approx\kappa_{B}^{2}\propto p^{2}$
(so $\rho=4\kappa_{\parallel}^{\prime}\kappa_{\perp}^{\prime}/r_{{\rm cr}}^{2}$,
where prime denotes $\partial_{p}$). Now we integrate this equation
and obtain the following spectrum at the shock

\begin{equation}
\bar{F}_{0}=\frac{C^{\prime}}{p^{3\sigma/\left(\sigma-1\right)}}\left[u_{1}+\sqrt{u_{1}^{2}+\rho p^{2}}\right]^{\frac{3\sigma}{2\left(\sigma-1\right)}}\left[u_{2}+\sqrt{u_{2}^{2}+\rho p^{2}}\right]^{\frac{3}{2\left(\sigma-1\right)}}\exp\left[-\frac{3}{2\Delta u}\left(\sqrt{u_{1}^{2}+\rho p^{2}}+\sqrt{u_{2}^{2}+\rho p^{2}}\right)\right]\label{eq:F0BarStandDSA}
\end{equation}
To compare the last result with the self-similar time-dependent solution
given by eqs.(\ref{eq:qOfpFinalGen}) and (\ref{eq:F0barGeneral}),
it is convenient to use the following dimensionless momentum

\[
\mathcal{P}=\sqrt{\rho}p/u_{2}=2^{3/2}\kappa_{B}/R_{s}u_{2}
\]
Note that the variable $v_{2}$, used in eq.(\ref{eq:vOfP}) can be
expressed through $\mathcal{P}$ as 
\begin{equation}
v_{2}=\frac{2^{1/4}}{\sqrt{\sigma\left(1-\beta\right)\mathcal{P}}},\label{eq:vVarForCompar}
\end{equation}
assuming the Bohm diffusion scaling for $\kappa\left(p\right)$ in
eq.(\ref{eq:qOfpFinalGen}). Again, it is not the purpose of this
comparison to obtain an exact position of the cut-off in the standard
DSA spectrum, but rather its momentum profile; so we consider $u_{1}$
and $R_{s}$ ``frozen'' and put $R_{s}\approx u_{1}t_{0}$. Meanwhile,
eq.(\ref{eq:F0BarStandDSA}) rewrites

\begin{equation}
\bar{F}_{0}=\frac{C}{\mathcal{P}^{3\sigma/\left(\sigma-1\right)}}\left[\sigma+\sqrt{\sigma^{2}+\mathcal{P}^{2}}\right]^{\frac{3\sigma}{2\left(\sigma-1\right)}}\left[1+\sqrt{1+\mathcal{P}^{2}}\right]^{\frac{3}{2\left(\sigma-1\right)}}\exp\left[-\frac{3}{2\left(\sigma-1\right)}\left(\sqrt{\sigma^{2}+\mathcal{P}^{2}}+\sqrt{1+\mathcal{P}^{2}}\right)\right]\label{eq:StandDSAdimLess}
\end{equation}
\begin{wrapfigure}{o}{0.5\columnwidth}%
\includegraphics[scale=0.47]{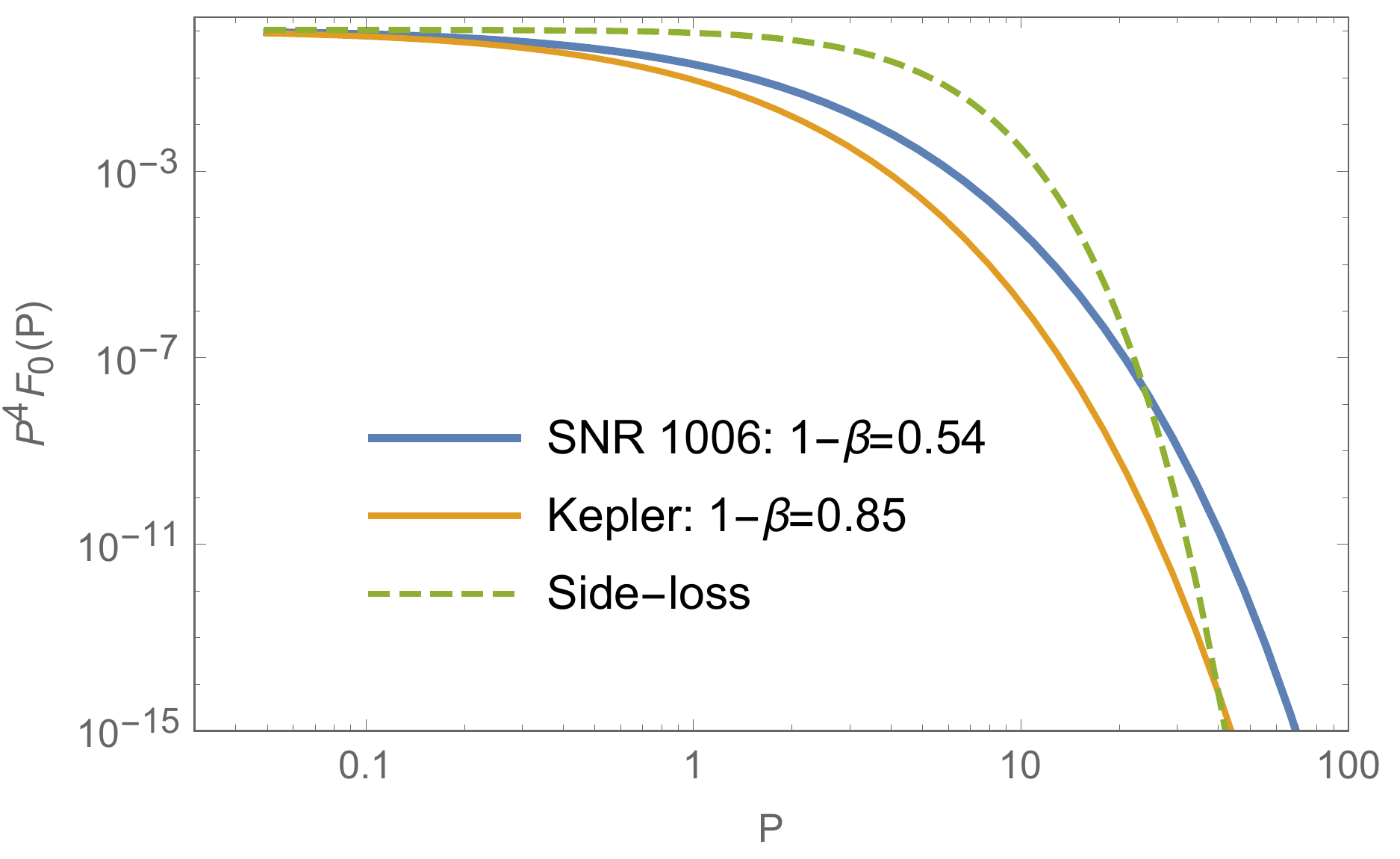}

\caption{Solid lines: loss-free spectrum from self-similar solution for indicated
values of expansion index $1-\beta$ and Bohm diffusion. Dashed line:
stationary solution with sideways losses. Both solutions are given
using normalized momentum $\mathcal{P}=2^{3/2}\kappa_{B}/R_{s}u_{2}$.
The standard DSA solution with particle losses implies that $\kappa_{\parallel}\kappa_{\perp}\propto\kappa_{B}^{2}\propto p^{2}$
(see text). The shock compression $\sigma=4$ and the spectra are
multiplied by $p^{4}.$ \label{fig:SpectraComparison}}
\end{wrapfigure}%
After adjusting the normalization factors in eqs.(\ref{eq:F0barGeneral})
and (\ref{eq:StandDSAdimLess}), both results are shown in Fig.\ref{fig:SpectraComparison},
extending the limiting case estimates above to a broader momentum
range. We have placed the DSA solutions with quite different mechanisms
of spectral steepening (particle escape vs shock expansion) on one
plot to emphasize the difference in their spectral shapes.

\begin{figure}
\includegraphics[scale=0.48]{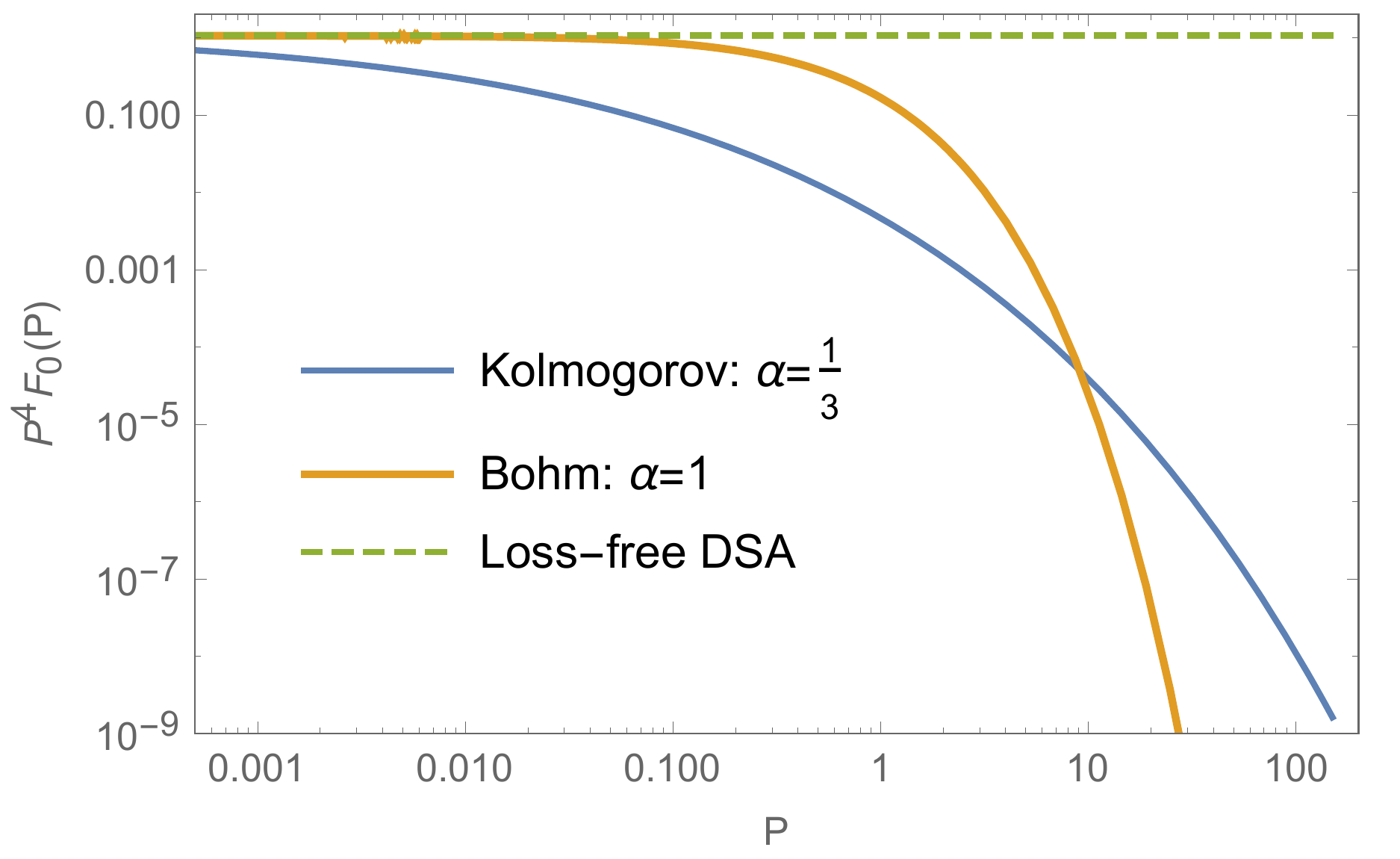}\includegraphics[scale=0.48]{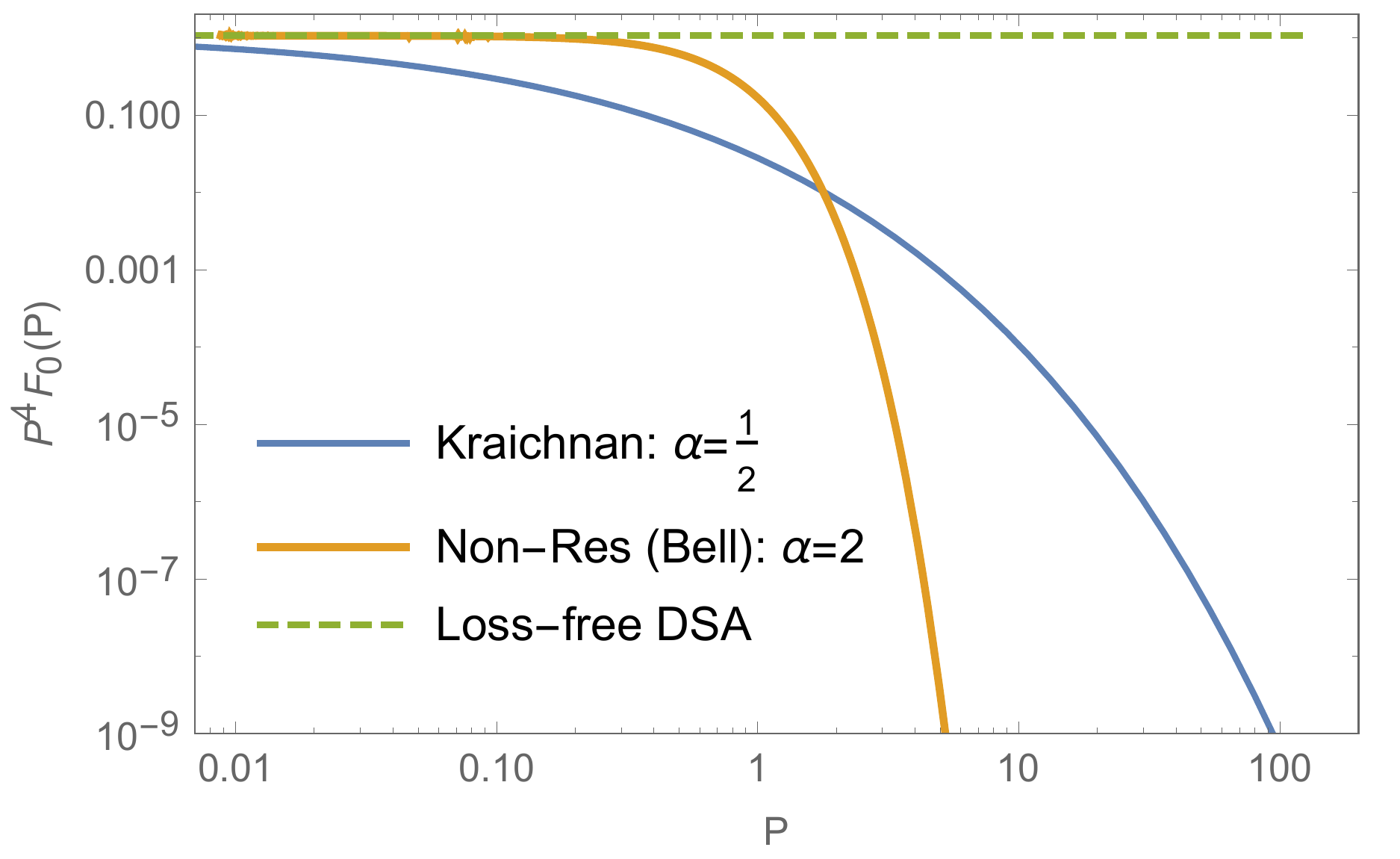}

\caption{The same as in Fig.\ref{fig:SpectraComparison}, but for different
turbulence models (solid lines) and for the expansion index $1-\beta=3/5$
(see Fig.\ref{fig:Shock-radius-(normalized}). The standard DSA solution
is shown for comparison on both panels (dashed lines) without cut-off.
\label{fig:SpectrumComp2}}
\end{figure}

\begin{figure}
\includegraphics[scale=0.48]{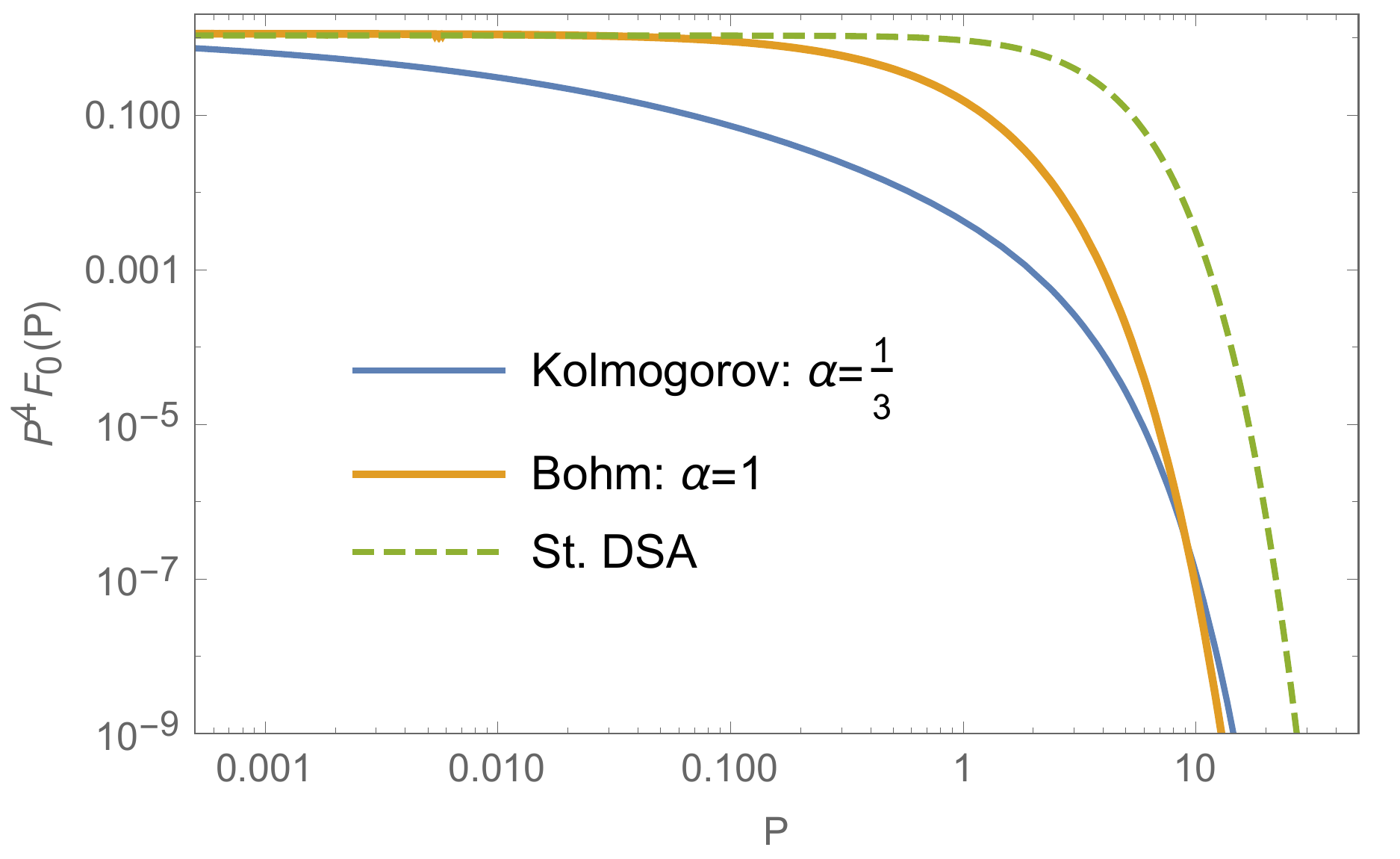}\includegraphics[scale=0.48]{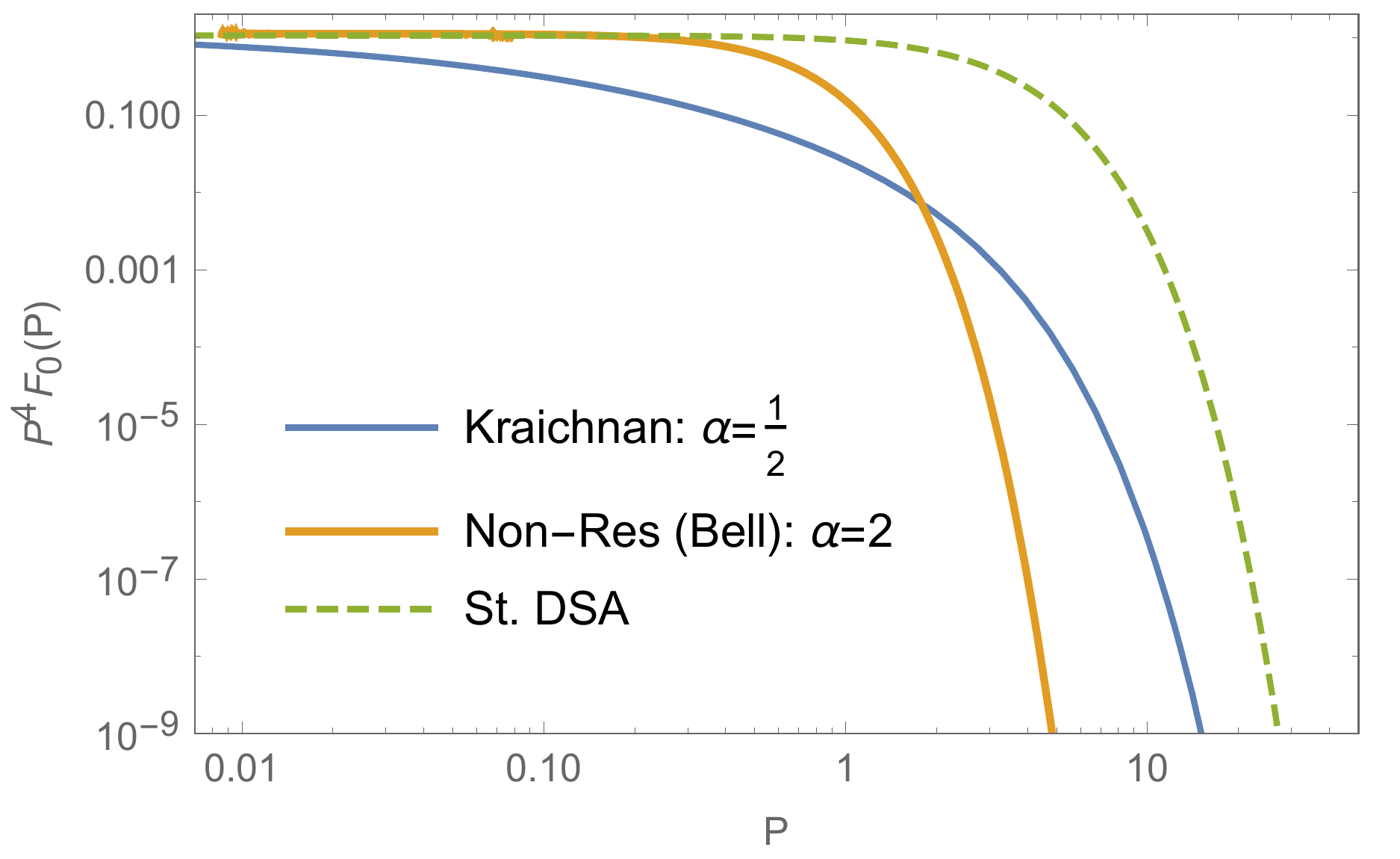}\caption{The same spectra as in Fig.\ref{fig:SpectrumComp2} but with cut-offs
imposed on all three spectra using the functional form obtained for
the standard DSA in eq.(\ref{eq:StandDSAdimLess}).\label{fig:FinalSpectraWCO}}
\end{figure}

The particle loss model employed above, shapes the form of the standard
DSA spectral cut-off through the product $\kappa_{\parallel}\kappa_{\perp}$,
which is firmly bound to $\kappa_{B}^{2}\left(p\right)$, according
to the widely accepted anisotropic diffusion paradigm, e.g., \citep{Drury83}.
This relation significantly constrains the standard DSA in explaining
the observed steep spectra. By contrast, the solution obtained in
this paper depends on $\kappa_{\parallel}\left(p\right)$, and in
the particular case of Bohm diffusion, it is characterized by a more
gradual steepening at high energies, Fig.\ref{fig:SpectraComparison}.
Only for $\kappa_{\parallel}\propto p^{2}$, the functional form of
the rollover is similar to the DSA cut-off (see below). Generally,
$\kappa_{\parallel}\left(p\right)$ strongly depends on the turbulence
model in use. We explore the potential of the present solution to
explain observed spectra by replacing the Bohm diffusion in Fig.\ref{fig:SpectraComparison}
with the three different turbulence models, often used in conjunction
with CR acceleration and propagation \citep{BlasiChemComp12,PtuskinPropRev12}.
These are the Kolmogorov model, with $\kappa\propto p^{1/3}$, Kraichnan
model with $\kappa\propto p^{1/2}$ and the short-scale non-resonant
turbulence model due to Bell instability with the diffusivity scaling
$\kappa\propto p^{2}.$ These three and the Bohm model are compared
in Fig.\ref{fig:SpectrumComp2} with a DSA result, first without cutoff.
Assuming then that the relation $\kappa_{\parallel}\kappa_{\perp}\approx\kappa_{B}^{2}$
holds up also at the edge of the acceleration zone, where the CR diffusion
is expected to be strongly anisotropic with $\kappa_{\parallel}\gg\kappa_{\perp}$,
we impose the DSA cut-off profile, eq.(\ref{eq:StandDSAdimLess}),
on the turbulent transport models listed above. The results are shown
in Fig.\ref{fig:FinalSpectraWCO} with a warning that, by contrast
to the spectra shown in Fig.\ref{fig:SpectrumComp2}, the solutions
with the imposed cut-offs are no longer the exact loss-free solutions
of eq.(\ref{eq:raveragedDC})

Although the mechanism of spectral steepening suggested in this paper
and the spectrum termination in the standard DSA invoke different
physical phenomena, the above comparison of the results may be helpful
in analyzing the emission spectra from bilateral SNRs and understanding
the production of bulk CR spectra. Specifically, if an observed spectrum
exhibits a gradual decline rather than a sharp one, it is more likely
that the steepening mechanisms considered in this paper are at work
and the cut-off is simply not reached yet. An improved approach would
be an incorporation of the lateral CR escape flux in our treatment
of Sec.\ref{subsec:Solution-of-Convection-Diffusion}. However, the
main difficulty that remains is an accurate calculation of the loss
function $\Lambda$. It requires a solution of a two-dimensional CR
transport problem at an interface between domains of strongly different
propagation regimes. Even if inside of the acceleration zone, $r_{\perp}<r_{{\rm cr}},$
$\left|z\right|\lesssim\kappa/u$ one may put $\kappa_{\parallel}\approx\kappa_{\perp}\approx\kappa_{B}\left(p\right),$
for $r>r_{{\rm cr}}$ the CR transport regime becomes strongly anisotropic
with $\kappa_{\parallel}\gg\kappa_{\perp}.$


\section{Summary \label{sec:Conclusions}}

In this paper, we have investigated the DSA operation in a spherical
SNR shock expanding in a homogeneous magnetic field. We have obtained
an exact self-similar solution for an arbitrary shock expansion index,
$1-\beta$, (defined as $R_{SNR}\propto t^{1-\beta}$), and particle
diffusivity $\kappa\left(p\right)$. Particle injection was assumed
to be efficient on two polar caps where the shock normal is inline
with the magnetic field within $\simeq45^{\circ}.$ Possible losses
of particles from the acceleration regions were neglected, except
for a comparison with the conventional DSA solution. The salient features
of the new solution are:
\begin{itemize}
\item particle momentum spectrum at the shock depends on the ratio of acceleration
time, $\tau_{a}\sim\kappa\left(p\right)/U_{{\rm sh}}^{2}$ to the
SNR dynamic time $t_{0}$ which, in turn, is related to the shock
expansion as $R_{SNR}\left(t\right)=R_{ST}\left(t/t_{0}\right)^{1-\beta}$,
or $U_{{\rm sh}}\left(t\right)=u_{1}\left(t_{0}/t\right)^{\beta}$(sec.\ref{subsec:SNR stage choice})
\begin{itemize}
\item for $\tau_{a}\ll t_{0},$ the spectrum has the standard DSA index
determined by the shock compression, $\sigma$, that is $f\propto p^{-q},$
with $q=3\sigma/\left(\sigma-1\right)$
\item the spectral index increases at larger $\tau_{a}$, and for $\tau_{a}\gg t_{0}$
it is suppressed by an exponential factor $\exp\left[-C\sqrt{\tau_{a}\left(p\right)/t_{0}}\right]$,
where $C\left(\sigma\right)\approx4.5$, eq.(\ref{eq:SpectrumLargeP})
\item this factor decreases with momentum much slower than the conventional
DSA cutoff, obtained, e.g., for the cross-field losses from the acceleration
zone of size $r_{{\rm cr}}$: $\propto\exp\left[-6\kappa_{B}\left(p\right)/\Delta ur_{{\rm cr}}\right]$,
eq.(\ref{eq:StandDSAdimLess})
\end{itemize}
\item the spatial distributions of particles upstream and downstream are
significantly different only at lower energies, with the downstream
distribution being broader, as in the planar shock solutions, Fig.\ref{fig:Spatial-profiles-of}
\begin{itemize}
\item at higher energies ($\tau_{a}\left(p\right)\gg t_{0})$ the upstream
and downstream acceleration zones are nearly equal, Fig.\ref{fig:Half-width-of-particle},
thus making a strong contrast to the plane-shock and spherically symmetric
(e.g., \citet{Kang2013ApJ}) solutions
\end{itemize}
\end{itemize}

\section{Discussion\label{sec:Discussion}}

There may be several reasons why the CR spectrum predicted by the
DSA disagrees with SNR observations and with the spectra of CRs arriving
at the Earth's atmosphere. We have shown that relaxing the shock stationarity
and spherical symmetry assumptions, typical of DSA models, alters
the following two DSA predictions. First, the SNR evolution affects
the acceleration stronger than in a spherically symmetric case because
the active acceleration zone on the shock grows in time while not
covering the entire shock surface. The growth time is the SNR dynamic
time $t_{0}$ and it affects the spectrum even well below the cut-off.
The latter is generally determined by the total duration of acceleration,
$t_{SNR}$, shock size, and the level of magnetic field amplification.
Of course, these three factors will set the ultimate cut-off energy
also for the present solution. Below that, the spectrum rolls over
when the particle acceleration time to momentum $p$, $\tau_{a}\left(p\right)\sim\kappa\left(p\right)/U_{sh}^{2}$,
becomes comparable with $t_{0}$. In historical SNRs from Table \ref{tab:Expansion-indices-and},
$t_{0}>t_{SNR}$ except for the SNR 1006. Generally, the condition
$t_{0}<t_{SNR}$ must hold for the self-similar solution to establish
at high energies where $\tau_{a}\left(p\right)>t_{0}$. This condition
can be relaxed if the maximum momentum is limited on physical grounds
(e.g., losses) to $\tau\left(p_{\text{max}}\right)<t_{SNR}$. In this
paper we have imposed a cut-off, $p_{\text{max}},$ only for comparison
with the standard DSA solution.

The second modification to the DSA is that accelerated particles stay
closer to the shock than in the ``standard'' planar and spherically
symmetric solutions. This effect is related to the growing surface
of acceleration that restricts particle convection time downstream.
It is more pronounced for high-energy particles since they spent most
of the time near the shock to gain high energy. The difference can
be seen by comparing Figs.\ref{fig:Spatial-profiles-of} and \ref{fig:Half-width-of-particle}
with the numerical solutions for a spherically symmetric acceleration
performed earlier, e.g., by \citet{Kang2013ApJ}. This result may
have implications for the interpretation of thin x-ray filaments observed
in the SNR 1006 in particular \citep{Bamba05,LongRaymond03}.

Although the solution obtained in this paper strictly applies to the
bilateral shock morphology, effects of changing magnetic field inclination
may be significant for the DSA in SNRs expanding in more complicated
magnetic environments. The SNR RX J 1713, for example, does not belong
to the bilateral type but it shows a patchy pattern of particle acceleration
around its rim \citep{HESSRXJ17132018}. This may be an effect of
ambient gas density, but it may be produced by variable field inclination
as well. Young SNRs usually have a radial field at the shock front
\citep{Dickel1991,West2017,Vink2018}, but the radial field geometry
is most probably a result of particle acceleration rather than its
prerequisite. Such SNRs may have started accelerating particles from
either bilateral state or from multiple hot spots (where $\vartheta_{nB}\lesssim\pi/4$)
that merge later. While this scenario is just a speculation, it is
not obvious how else one can reconcile the radial field morphology
all around the rim with an arbitrary, possibly homogeneous initial
field into which the progenitor star has exploded. The SNR 1006 is,
perhaps, a telltale counterexample to the puzzling omni-radial field
geometry of young remnants.

\subsection{Acceleration Nonlinearity and Turbulent Heating}

The nonlinear shock modification, mentioned in the Introduction, is
often considered as irrelevant to the SNR shocks for the following
two reasons. First, it predicts flatter than the linear DSA spectra,
not steeper ones, as suggested by observations (see \citealp{MDru01}
for a review of nonlinear solutions). Second, it is not observed in
hybrid simulations. We believe that neither argument is convincing.
Starting from the apparent problem with the observations, our new
solution that steepens at high energies can easily compensate for
the nonlinear flattening. It is also impossible to ignore the fact
that the nonlinear DSA regime follows \emph{directly }from equations
describing the acceleration. The exact criteria for transitioning
of the acceleration to the nonlinear regime can be obtained from a
bifurcation analysis of the steady-state kinetic solution. This solution
links the problem parameters, discussed below, by placing them on
a surface in the multidimensional parameter space. The solution, however,
needs to be obtained in analytic form to understand the transition
since it is unstable on a certain part of that surface and thus inaccessible
for numerical or ``semi-analytic'' approaches.

The \emph{critical} parameters characterizing the transition are (i)
injection rate of thermal particles (roughly a fraction of accelerated
CRs, $\eta$), (ii) shock Mach number, $M$, (iii) maximum momentum,
$p_{{\rm max}}$, and (iv) heating rate of the upstream plasma. While
the parameters (i-iii) are usually reasonably known (with some reservations
about $p_{{\rm max}}$, discussed below), the heating rate is a highly
uncertain parameter. A zero-order approximation of the (i-iii) part
of the parameter space where the acceleration must be in an efficient
nonlinear regime is the following: $p_{\text{max}}/m_{p}c\gtrsim10^{3},$
$M\gtrsim20$, and the injection rate $\eta\gtrsim10^{-3}.$ Each
of these thresholds can be lowered somewhat at the expense of the
other two. However, the maximum momentum cannot be significantly decreased,
since the requirement for the nonlinear transition, which is in fact
$\ln\left(p_{\text{max}}/mc\right)\gg1$, simply follows from the
scaling of the CR pressure, $P_{CR}$, with $p_{\text{max}}$. In
a pre-transition test-particle acceleration regime, when the CR spectrum
is $\propto p^{-4}$, the CR pressure grows slowly with $p_{\text{max}},$
namely, $P_{CR}\propto\eta\ln\left(p_{\text{max}}/m_{p}c\right)$.
Since $P_{CR}$ must become comparable to the shock ram pressure for
the transition to occur, and since $\eta\ll1$, the required $p_{\text{max }}$
is quite high and not yet attainable in the current hybrid, let alone
PIC simulations. Not surprisingly, the nonlinear transition is not
observed in such simulations. Of course, it is inexorably present
to all numerical solutions satisfying the above set of criteria, e.g.,
\citep{KangJones2002}.

Meanwhile, upon transitioning to the nonlinear acceleration regime,
the spectrum becomes much flatter, $\propto p^{-3.5}$, so that the
CR pressure grows much sharper with $p_{\text{max}},$ $P_{CR}\propto p_{\text{max}}^{1/2}$.
It is this feedback loop connecting the CR pressure and the spectral
index that makes a hard (hysteresis type) transition to the nonlinear
acceleration regime. Under these circumstances, the heating must be
provided by a collisionless turbulence dissipation (via plasma microinstabilities)
in the upstream flow permeated by the CRs. They drive strong turbulence,
and a complicated coupling to the driver (CRs) exacerbates the problem
of turbulence dissipation. There are no analytic studies that quantitatively
address this problem. Since scales associated with large gyroradii
of particles at highest energies are immensely disparate, the problem
is not tractable numerically either. Therefore, the heating rate of
the upstream plasma remains largely unknown.

Not having a reliable heating rate at hand, we can still parameterize
it and obtain the nonlinear solution for the DSA. This solution, shown
in Fig.6 of \citep{MDru01}, proves very sensitive to that same heating
rate. For example, if we decrease this parameter by about two orders
of magnitude across its high-sensitivity range, the spectrum will
change from a pure test particle, $f\propto p^{-4}$, to strongly
nonlinear, $f\propto p^{-3.5}$. Note that these are the limiting
cases with vastly different acceleration efficiencies. In the latter
case, most of the particle pressure is deposited near $p_{{\rm max}}$,
which appears to be inconsistent with the continuous spectrum steepening
obtained in this paper. Indeed, it must diminish the nonlinear effects
by restricting the growth of an \emph{effective }$p_{\text{max}}$
that controls the nonlinear transition. However, appealing to our
discussion of the spectrum steepening illustrated by Fig.\ref{fig:Spherical-shock-surface}
we can reconcile the opposite trends. First of all, the nonlinear
transition must begin in the area (near magnetic axis) with the longest
acceleration history where the local value of $p_{\text{max}}$ is
higher. In the outskirts of the acceleration zone, recently injected
particles with much lower $p_{\text{max}}$ will dominate the spectrum.
As we already argued earlier, the line-of-sight- or acceleration-zone-averaged
spectrum must appear as a superposition of these two extremes. Therefore,
the nonlinear hardening of the central spectrum will be masked by
the softer periphery.

Apart from the above-discussed variation of the spectrum along the
shock surface, an intrinsic time dependence of nonlinear acceleration
should also play an important role in establishing the integrated
spectrum. This variability occurs independently of the spherical shock
effects, addressed in this paper. As we argued above, the nonlinear
transition to an efficient acceleration regime is quasi-abrupt in
character. It occurs when $p_{\text{max}}\left(t\right)$ surpasses
its critical value. Soon thereafter, an increased CR pressure reacts
back on the shock structure thereby decreasing the injection and increasing
heating rate, thus driving the system back to its subcritical state.
The backtransition is, however, not sustainable since $p_{\text{max}}$
continues to grow, so the next forward transition must follow. These
limit-cycle oscillations have been \emph{predicted }from the analytic
solution of the nonlinear acceleration problem. Indeed, they appear
to be present in numerical solutions due to \citet{KangJones2002}
(see their Fig.7 and a recent discussion of the phenomenon by \citealp{MALKOV201820}).
As in the case of spatial variation of acceleration efficiency, the
limit-cycle oscillations must result in a spectral index between its
efficient and test-particle values. Since the steepening of the test-particle
solution obtained in the present paper is significant, the net result
may be in the ballpark of the current observations \citep{Aharon2018}.
Again, the quantitative predictions require a resolution of complicated
issues associated with particle losses and turbulent heating efficiency.

\subsection{Concluding Remarks}

With all the above limitations in mind, we may still ask how important
for the integrated CR spectrum the obtained solution might be. To
answer this question, in addition to more realistic particle losses
discussed in the Introduction, one needs to self-consistently include
the particle injection rate, $Q$, and, most importantly, their diffusion
coefficient $\kappa\left(p\right)$ into the obtained spectrum. A
promising strategy is to extract these acceleration parameters from
the hybrid (and therefore \emph{very }limited-time) shock simulation,
taken at a \emph{few }short SNR evolution periods, and interpolate
them in the solution obtained above. Using a similar approach \citet{Hanusch2019ApJ}
recently provided a viable explanation of the anomaly of proton/Helium
ratio rigidity spectrum, presumably accumulated over the free-expansion
to Sedov-Taylor SNR stages. It is important to understand, that the
p/He \emph{ratio} rigidity spectrum is not changed by the spectral
hardening or softening described above, or by any other electromagnetic
processes, such as propagation in, or escape from, the galaxy.

Returning to the spectrum softening mechanism suggested in this paper
and looking at Fig.\ref{fig:SpectrumComp2} one may come to a disappointing
conclusion. By adjusting the CR diffusion model for $\kappa\left(p\right)$
within a range between Kolmogorov and non-resonant small-scale (Bell)
turbulence one can fit almost any observed spectrum. On the bright
side, we obtained liberation from the notoriously inflexible linear
DSA spectrum that is increasingly in conflict with the rapidly improving
observations. Moreover, while the soft $\gamma$-ray spectra of young
SNR can be explained within the standard (linear or non-linear) DSA
framework only by introducing \textquotedbl early\textquotedbl{}
cutoffs\footnote{There are alternative spectral softening mechanisms, e.g. \citep{MD06},
not discussed here because they require non-diffusive particle transport,
thus departing from the convection-diffusion equation \citep{KirkBraidedMF96,Zimbardo2013ApJ}
(see, however, conclusions in \citep{malkov2017exact} regarding some
issues with the justification of non-diffusive CR transport).}, Fig.\ref{fig:SpectrumComp2} shows that the approach described in
this paper may explain them without a commonly assumed and in practice
inevitably parameterized exponential cutoff! The $\kappa$ -dependence
on energy naturally produces a more gradual decay of the spectrum.
This implies, in particular, that the steep spectra of SN 1006, Tycho,
Cas A, RX J 1713 (see \citealp{Aharon2018} for further discussion
and references), as reported in the multi-TeV energy band, cannot
rule out these objects as PeVatrons. Whether the highest-energy particles
in these SNRs can reach the PeV range in detectable numbers, is a
different issue falling outside the scope of this paper.

\acknowledgements{}

Work of MM is supported by NASA Astrophysics Theory Program under
grant 80NSSC17K0255.

\pagebreak{}

\bibliographystyle{aasjournal}
\bibliography{}

\begin{thebibliography}{}
\expandafter\ifx\csname natexlab\endcsname\relax\def\natexlab#1{#1}\fi
\providecommand{\url}[1]{\href{#1}{#1}}
\providecommand{\dodoi}[1]{doi:~\href{http://doi.org/#1}{\nolinkurl{#1}}}
\providecommand{\doeprint}[1]{\href{http://ascl.net/#1}{\nolinkurl{http://ascl.net/#1}}}
\providecommand{\doarXiv}[1]{\href{https://arxiv.org/abs/#1}{\nolinkurl{https://arxiv.org/abs/#1}}}

\bibitem[{{Aharonian} {et~al.}(2018){Aharonian}, {Yang}, \& {de O{\~n}a
  Wilhelmi}}]{Aharon2018}
{Aharonian}, F., {Yang}, R., \& {de O{\~n}a Wilhelmi}, E. 2018, ArXiv e-prints.
\newblock \doarXiv{1804.02331}

\bibitem[{{Axford}(1981)}]{Axford1981_ICRC}
{Axford}, W.~I. 1981, International Cosmic Ray Conference, 12, 155

\bibitem[{{Bamba} {et~al.}(2005){Bamba}, {Yamazaki}, {Yoshida}, {Terasawa}, \&
  {Koyama}}]{Bamba05}
{Bamba}, A., {Yamazaki}, R., {Yoshida}, T., {Terasawa}, T., \& {Koyama}, K.
  2005, \apj, 621, 793, \dodoi{10.1086/427620}

\bibitem[{{Bateman}(1955)}]{Bateman1955}
{Bateman}, H. 1955, {Higher transcendental functions}

\bibitem[{{Bell} {et~al.}(2019){Bell}, {Matthews}, \&
{Blundell}}]{Bell2019}
{Bell}, A.~R., {Matthews}, J.~H., \& {Blundell}, K.~M. 2019,
arxiv: 1906.12240, to appear in MNRAS

\bibitem[{{Berezhko} {et~al.}(1994){Berezhko}, {Yelshin}, \&
  {Ksenofontov}}]{Ber94DruSuppr}
{Berezhko}, E.~G., {Yelshin}, V.~K., \& {Ksenofontov}, L.~T. 1994,
  Astroparticle Physics, 2, 215, \dodoi{10.1016/0927-6505(94)90043-4}

\bibitem[{{Bisnovatyi-Kogan} \& {Silich}(1995)}]{Bisnovatyi95}
{Bisnovatyi-Kogan}, G.~S., \& {Silich}, S.~A. 1995, Reviews of Modern Physics,
  67, 661, \dodoi{10.1103/RevModPhys.67.661}

\bibitem[{{Blandford} \& {Eichler}(1987)}]{BlandEich87}
{Blandford}, R., \& {Eichler}, D. 1987, \physrep, 154, 1

\bibitem[{Blasi \& Amato(2012)}]{BlasiChemComp12}
Blasi, P., \& Amato, E. 2012, Journal of Cosmology and Astroparticle Physics,
  2012, 010

\bibitem[{{Caprioli} \& {Spitkovsky}(2014)}]{CaprSpitk14a}
{Caprioli}, D., \& {Spitkovsky}, A. 2014, \apj, 783, 91,
  \dodoi{10.1088/0004-637X/783/2/91}

\bibitem[{{Cassam-Chena{\"i}} {et~al.}(2008){Cassam-Chena{\"i}}, {Hughes},
  {Reynoso}, {Badenes}, \& {Moffett}}]{Cassam-Ch2008}
{Cassam-Chena{\"i}}, G., {Hughes}, J.~P., {Reynoso}, E.~M., {Badenes}, C., \&
  {Moffett}, D. 2008, \apj, 680, 1180, \dodoi{10.1086/588015}

\bibitem[{{Chevalier}(1977)}]{Chevalier1977ARA&A}
{Chevalier}, R.~A. 1977, \araa, 15, 175,
  \dodoi{10.1146/annurev.aa.15.090177.001135}

\bibitem[{{D'Angelo} {et~al.}(2016){D'Angelo}, {Blasi}, \&
  {Amato}}]{DAngeloBlasi2016}
{D'Angelo}, M., {Blasi}, P., \& {Amato}, E. 2016, \prd, 94, 083003,
  \dodoi{10.1103/PhysRevD.94.083003}

\bibitem[{{Dickel} {et~al.}(1991){Dickel}, {van Breugel}, \&
  {Strom}}]{Dickel1991}
{Dickel}, J.~R., {van Breugel}, W.~J.~M., \& {Strom}, R.~G. 1991, \aj, 101,
  2151, \dodoi{10.1086/115837}

\bibitem[{{Drury}(1983)}]{Drury83}
{Drury}, L.~O. 1983, Reports on Progress in Physics, 46, 973

\bibitem[{{Drury}(2011)}]{DruryEscape11}
---. 2011, \mnras, 415, 1807, \dodoi{10.1111/j.1365-2966.2011.18824.x}

\bibitem[{{Ellison} {et~al.}(1995){Ellison}, {Baring}, \&
  {Jones}}]{Ellison1995ApJ}
{Ellison}, D.~C., {Baring}, M.~G., \& {Jones}, F.~C. 1995, \apj, 453, 873,
  \dodoi{10.1086/176447}

\bibitem[{{Gabici} {et~al.}(2019){Gabici}, {Evoli}, {Gaggero}, {Lipari},
  {Mertsch}, {Orlando}, {Strong}, \& {Vittino}}]{Gabici2019}
{Gabici}, S., {Evoli}, C., {Gaggero}, D., {et~al.} 2019, arXiv e-prints,
  arXiv:1903.11584.
\newblock \doarXiv{1903.11584}

\bibitem[{{H.~E.~S.~S.~Collaboration}
  {et~al.}(2018){H.~E.~S.~S.~Collaboration}, {Abdalla}, {Abramowski},
  {Aharonian}, {Benkhali}, {Akhperjanian}, {Andersson}, {Ang{\"u}ner},
  {Arrieta}, {Aubert}, \& et~al.}]{HESSRXJ17132018}
{H.~E.~S.~S.~Collaboration}, {Abdalla}, H., {Abramowski}, A., {et~al.} 2018,
  \aap, 612, A6, \dodoi{10.1051/0004-6361/201629790}

\bibitem[{{Hanusch} {et~al.}(2019){Hanusch}, {Liseykina}, \&
  {Malkov}}]{Hanusch2019ApJ}
{Hanusch}, A., {Liseykina}, T.~V., \& {Malkov}, M. 2019, \apj, 872, 108,
  \dodoi{10.3847/1538-4357/aafdae}

\bibitem[{{Kang} \& {Jones}(2002)}]{KangJones2002}
{Kang}, H., \& {Jones}, T.~W. 2002, Journal of Korean Astronomical Society, 35,
  159, \dodoi{10.5303/JKAS.2002.35.4.159}

\bibitem[{{Kang} {et~al.}(2013){Kang}, {Jones}, \& {Edmon}}]{Kang2013ApJ}
{Kang}, H., {Jones}, T.~W., \& {Edmon}, P.~P. 2013, \apj, 777, 25,
  \dodoi{10.1088/0004-637X/777/1/25}

\bibitem[{{Kirk} {et~al.}(1996){Kirk}, {Duffy}, \& {Gallant}}]{KirkBraidedMF96}
{Kirk}, J.~G., {Duffy}, P., \& {Gallant}, Y.~A. 1996, \aap, 314, 1010

\bibitem[{{Long} {et~al.}(2003){Long}, {Reynolds}, {Raymond}, {Winkler},
  {Dyer}, \& {Petre}}]{LongRaymond03}
{Long}, K.~S., {Reynolds}, S.~P., {Raymond}, J.~C., {et~al.} 2003, \apj, 586,
  1162, \dodoi{10.1086/367832}

\bibitem[{Malkov(2018)}]{MALKOV201820}
Malkov, M. 2018, Nuclear and Particle Physics Proceedings, 297-299, 20 ,
  \dodoi{https://doi.org/10.1016/j.nuclphysbps.2018.07.004}

\bibitem[{Malkov(2017)}]{malkov2017exact}
Malkov, M.~A. 2017, Physical Review D, 95, 023007

\bibitem[{{Malkov} \& {Diamond}(2006)}]{MD06}
{Malkov}, M.~A., \& {Diamond}, P.~H. 2006, \apj, 642, 244,
  \dodoi{10.1086/430344}

\bibitem[{{Malkov} {et~al.}(2013){Malkov}, {Diamond}, {Sagdeev}, {Aharonian},
  \& {Moskalenko}}]{MetalEsc13}
{Malkov}, M.~A., {Diamond}, P.~H., {Sagdeev}, R.~Z., {Aharonian}, F.~A., \&
  {Moskalenko}, I.~V. 2013, \apj, 768, 73, \dodoi{10.1088/0004-637X/768/1/73}

\bibitem[{{Malkov} \& {Drury}(2001)}]{MDru01}
{Malkov}, M.~A., \& {Drury}, L.~O. 2001, Reports on Progress in Physics, 64,
  429

\bibitem[{{Malkov} \& {V\"olk}(1995)}]{mv95}
{Malkov}, M.~A., \& {V\"olk}, H.~J. 1995, \aap, 300, 605

\bibitem[{{McKee} \& {Truelove}(1995)}]{McKeeTruelove95}
{McKee}, C.~F., \& {Truelove}, J.~K. 1995, \physrep, 256, 157,
  \dodoi{10.1016/0370-1573(94)00106-D}

\bibitem[{{Nava} \& {Gabici}(2013)}]{GabiciAnisEsc12}
{Nava}, L., \& {Gabici}, S. 2013, \mnras, 355, \dodoi{10.1093/mnras/sts450}

\bibitem[{{Pais} {et~al.}(2018){Pais}, {Pfrommer}, {Ehlert}, \&
  {Pakmor}}]{Pfrommer2018MNRAS}
{Pais}, M., {Pfrommer}, C., {Ehlert}, K., \& {Pakmor}, R. 2018, \mnras, 478,
  5278, \dodoi{10.1093/mnras/sty1410}

\bibitem[{{Prishchep} \& {Ptuskin}(1981)}]{PtusPrish1981}
{Prishchep}, V.~L., \& {Ptuskin}, V.~S. 1981, \azh, 58, 779

\bibitem[{{Ptuskin}(2012)}]{PtuskinPropRev12}
{Ptuskin}, V. 2012, Astroparticle Physics, 39, 44,
  \dodoi{10.1016/j.astropartphys.2011.11.004}

\bibitem[{Thomas \& Winske(1990)}]{thomas1990two}
Thomas, V., \& Winske, D. 1990, Geophysical Research Letters, 17, 1247

\bibitem[{{Toptygin}(1980)}]{Toptygin80}
{Toptygin}, I.~N. 1980, Space Science Reviews, 26, 157

\bibitem[{{Vink} \& {Zhou}(2018)}]{Vink2018}
{Vink}, J., \& {Zhou}, P. 2018, Galaxies, 6, 46,
  \dodoi{10.3390/galaxies6020046}

\bibitem[{{V{\"o}lk} {et~al.}(2003){V{\"o}lk}, {Berezhko}, \&
  {Ksenofontov}}]{VoelkInj03}
{V{\"o}lk}, H.~J., {Berezhko}, E.~G., \& {Ksenofontov}, L.~T. 2003, \aap, 409,
  563, \dodoi{10.1051/0004-6361:20031082}

\bibitem[{{West} {et~al.}(2017){West}, {Jaffe}, {Ferrand}, {Safi-Harb}, \&
  {Gaensler}}]{West2017}
{West}, J.~L., {Jaffe}, T., {Ferrand}, G., {Safi-Harb}, S., \& {Gaensler},
  B.~M. 2017, \apjl, 849, L22, \dodoi{10.3847/2041-8213/aa94c4}

\bibitem[{{Zimbardo} \& {Perri}(2013)}]{Zimbardo2013ApJ}
{Zimbardo}, G., \& {Perri}, S. 2013, \apj, 778, 35,
  \dodoi{10.1088/0004-637X/778/1/35}

\end{thebibliography}

\end{document}